%% file: main.tex
\documentclass{LMCS}
\usepackage{code}
\usepackage{ifthen}
\usepackage{enumerate}
\usepackage{latexsym,amssymb}
\usepackage{amsthm}
\usepackage{stmaryrd}
\usepackage{mathpartir}
\usepackage{unlist}
\usepackage{alltt}
\usepackage{verbatim} % for comment environment
\usepackage{proof}
\usepackage{hyperref}

\newcounter{newrule}
\input{defs-stepless}
\input{defs2-stepless}

\def\doi{7 (2:16) 2011}
\lmcsheading%
{\doi}
{1--37}
{}
{}
{Jan.~\phantom08, 2010}
{Jun.~\phantom07, 2011}
{}

\begin{document}
\title{Logical Step-Indexed Logical Relations\rsuper*}

\author[D. Dreyer]{Derek Dreyer\rsuper a}
\address{{\lsuper a}MPI-SWS, Germany}
\email{dreyer@mpi-sws.org}

\author[A. Ahmed]{Amal Ahmed\rsuper b}
\address{{\lsuper b}Indiana University, USA}
\email{amal@cs.indiana.edu}

\author[L. Birkedal]{Lars Birkedal\rsuper c}
\address{{\lsuper c}IT University of Copenhagen, Denmark}
\email{birkedal@itu.dk}

\titlecomment{{\lsuper*}This is an expanded and revised version of a paper that
  appeared at LICS'09.  In addition to presenting improved results,
  this version corrects a technical flaw in the earlier paper (see
  Section~\ref{sec:lics-comparison}).}

%\date{}
% \thispagestyle{empty}

%\vspace{-0.4in}

\begin{abstract}

\input{abstract}

\end{abstract}

\subjclass{??}

\maketitle

\input{intro}

\input{lang}

\input{logic}

\input{logrel}

\input{examples}

\input{discussion}

\input{lics-comparison}

\input{related}

{
\footnotesize
 \bibliographystyle{plain}
 \bibliography{main}
}

\clearpage
\appendix
\input{appendix}
\input{lang-stepless}

\end{document}

%% file: defs-stepless.tex
\newcommand{\omitthis}[1]{}
\newcommand{\presec}{}
\newcommand{\postsec}{}

\newcommand{\prepara}{}
\newcommand{\predisp}{}
\newcommand{\postdisp}{}
\newcommand{\mit}[1]{\mathit{#1}}
\newcommand{\mrm}[1]{\mathrm{#1}}
\newcommand{\mtt}[1]{\mathtt{#1}}
\newcommand{\mcal}[1]{\mathcal{#1}}

\newcommand{\msf}[1]{\mathsf{#1}}

\newenvironment{nop}{}{}

\newenvironment{sdisplaymath}{
\begin{nop}\small\begin{displaymath}}{
\end{displaymath}\end{nop}\ignorespacesafterend}

\newenvironment{smathpar}{
\begin{nop}\small\begin{mathpar}}{
\end{mathpar}\end{nop}\ignorespacesafterend}

\newenvironment{stackAux}[2]{%
\setlength{\arraycolsep}{0pt}
\begin{array}[#1]{#2}}{
\end{array}}

\newenvironment{stackTL}{
\begin{stackAux}{t}{l}}{\end{stackAux}}

\newtheorem{theoremX}{Theorem}[section]

\newtheorem{lemmaX}[theoremX]{Lemma}

\newtheorem{conjectureX}[theoremX]{Conjecture}

\newtheorem{corollaryX}[theoremX]{Corollary}

\theoremstyle{definition}
\newtheorem{definitionX}[theoremX]{Definition}

\renewenvironment{cases}{\begin{enumerate}[\hbox to8 pt{\hfill}]}{\end{enumerate}}

\newcommand{\casen}[1]{\item\item\noindent{\hskip-12 pt\bf Case #1:}\ }

\newcommand{\dom}{\mrm{dom}}

\newcommand{\fv}{\mathrm{FV}}
\newcommand{\FV}{\mathrm{FV}}

\newcommand{\subst}[3]{{#3}[{#1}/{#2}]} 
 
\newcommand{\defeq}{\stackrel{\mathrm{def}}{=}}

\newcommand{\lsem}{\left\llbracket}
\newcommand{\rsem}{\right\rrbracket}
\newcommand{\sembrace}[1]{\lsem{#1}\rsem}

\newcommand{\empctx}{\cdot}

\newcommand{\seq}[1]{\overline{#1}}
\newcommand{\val}{\mrm{val}}

\newcommand{\bnfalt}{{\bf \,\,\mid\,\,}}
\newcommand{\bnfdef}{{\bf ::=}}

\newcommand{\fref}{\mathsf{F}^{\mu!}}
\newcommand{\fmu}{\mathsf{F}^{\mu}}

\newcommand{\twf}[2]{{#1} \vdash {#2}}
\newcommand{\twfok}[2]{\FV(#2) \subseteq {#1}}
\newcommand{\ttwf}[2]{{#1} \vdash {#2}}
\newcommand{\judg}[3]{{#1} \vdash {#2} : {#3}}

\newcommand{\tyfont}[1]{\msf{#1}} 
\newcommand{\tunit}{\tyfont{unit}}
\newcommand{\tbool}{\tyfont{bool}}
\newcommand{\tint}{\tyfont{int}}

\newcommand{\tpair}[2]{#1 \times #2}
\newcommand{\tsum}[2]{#1 + #2}
\newcommand{\tfun}[2]{{#1} \rightarrow {#2}}

\newcommand{\trec}[2]{\mu {#1}.\,#2}
\newcommand{\tall}[2]{\forall #1.\,#2}
\newcommand{\texist}[2]{\exists #1.\,#2}

\newcommand{\tmfont}[1]{\mtt{#1}}

\newcommand{\op}{\mathit{o}}

\newcommand{\eunit}{\tmfont{\langle\rangle}}
\newcommand{\etrue}{\tmfont{true}}
\newcommand{\efalse}{\tmfont{false}}
\newcommand{\eif}[3]{\tmfont{if}\,#1\,\tmfont{then}\,#2\,\tmfont{else}\,#3}
\newcommand{\etuple}[1]{\langle{#1}\rangle}
\newcommand{\epair}[2]{\etuple{{#1},{#2}}}

\newcommand{\efst}[1]{\tmfont{fst}\,#1}
\newcommand{\esnd}[1]{\tmfont{snd}\,#1}
\newcommand{\efix}[3]{\mathcd{fix}\,#1(#2).\,#3}
\newcommand{\efun}[3]{\lambda#1{\,:\,}#2.\,#3}

\newcommand{\eunf}[1]{\lambda#1.}
\newcommand{\eapp}[2]{#1\,#2}

\newcommand{\efold}[1]{\tmfont{roll}\,{#1}}
\newcommand{\eroll}[1]{\tmfont{roll}\,{#1}}
\newcommand{\efoldty}[2]{\tmfont{roll}_{#1}\,{#2}}

\newcommand{\eunfold}[1]{\tmfont{unroll}\,#1}
\newcommand{\eunroll}[1]{\tmfont{unroll}\,#1}
\newcommand{\einl}[1]{\tmfont{inl}\,{#1}}
\newcommand{\einlty}[2]{\tmfont{inl}_{#1}\,{#2}}
\newcommand{\einr}[1]{\tmfont{inr}\,{#1}}
\newcommand{\einrty}[2]{\tmfont{inr}_{#1}\,{#2}}
\newcommand{\ecase}[5]{\tmfont{case}\,{#1}\,\tmfont{of}\,\tmfont{inl}\,{#2}\!\Rightarrow\!{#3}\,{\mid}\,\tmfont{inr}\,{#4}\!\Rightarrow\!{#5}}

\newcommand{\ecasebrneg}[5]{\begin{stackTL}\tmfont{case}\,{#1}\,\tmfont{of}\,\tmfont{inl}\,{#2}\!\Rightarrow\!{#3}\\\quad\mid\!\tmfont{inr}\,{#4}\!\Rightarrow\!{#5}\end{stackTL}}

\newcommand{\etabs}[2]{\Lambda #1.\,#2}
\newcommand{\etapp}[2]{#1\,#2}
\newcommand{\epack}[3]{\tmfont{pack}\,#1,#2\,\tmfont{as}\,#3}

\newcommand{\eunpack}[4]{\tmfont{unpack}\,{#1}\,\tmfont{as}\,{#2,#3}\,\tmfont{in}\,{#4}}

\newcommand{\step}{\leadsto}
\newcommand{\stepzero}{\step^0}
\newcommand{\stepone}{\step^1}
\newcommand{\stepmany}{\step^*}
\newcommand{\stepk}[1]{\stackrel{#1}{\step}}
\newcommand{\evalsto}{\step^*}
\newcommand{\termin}[1]{#1 \Downarrow}

\newcommand{\ciuleq}{\preceq}
\newcommand{\ciuleqo}{\ciuleq_1}
\newcommand{\ciuleqt}{\ciuleq_2}

\newcommand{\holewith}[1]{[#1]}
\newcommand{\hw}[1]{\holewith{#1}}
\newcommand{\hole}{\holewith{\cdot}}
\newcommand{\ctxt}{C}
\newcommand{\ectxt}{E}
\newcommand{\ctxarrow}{\rightsquigarrow}
\newcommand{\ctxaprx}{\preceq^{\mit{ctx}}}
\newcommand{\logaprx}{\preceq^{\mit{log}}}

\newcommand{\ctxeqv}{\approx^{\mit{ctx}}}
\newcommand{\ciueqv}{\approx^{\mit{ciu}}}

\newcommand{\pextends}[1]{\sqsupseteq}

\newcommand{\Vrel}[1]{\mcal{V}\sembrace{#1}}
\newcommand{\Vrelsym}[1]{\mcal{V}^{\approx\!}\sembrace{#1}}

\newcommand{\Crel}[1]{\mcal{E}\sembrace{#1}}
\newcommand{\Crelsym}[1]{\mcal{E}^{\approx\!}\sembrace{#1}}

\newcommand{\thelogic}{LSLR}

\newcommand{\Rel}{\mathrm{Rel}}

\newcommand{\gra}{\con}
\newcommand{\gr}{\vcon;\rcon}
\newcommand{\ts}{\vdash}

\newcommand{\ciuaprx}{\preceq^{\mit{ciu}}}

\newcommand{\limp}{\Rightarrow}

\newcommand{\later}{{\triangleright}}
\newcommand{\lift}{\later}

\newcommand{\interp}[3]{\sembrace{#1}\!{#3}{#2}}
\newcommand{\atominterp}[1]{\mathcal{I}(#1)}

\newcommand{\myinfer}[2]{\infer{#1}{#2}}

\newcommand{\vrel}[2]{\mathrm{VRel}(#1,#2)}
\newcommand{\trel}[2]{\mathrm{TRel}(#1,#2)}

\newcommand{\vcon}{\mathcal{X}}
\newcommand{\rcon}{\mathcal{R}}
\newcommand{\acon}{\mathcal{P}}
\newcommand{\pcon}{\mathcal{P}}

\newcommand{\con}{\mathcal{C}}

\newcommand{\Prop}{\mathrm{Prop}}

\newcommand{\OVal}{\mathrm{Val}}
\newcommand{\VRel}[2]{\mathrm{VRel}(#1,#2)}
\newcommand{\valof}{\downarrow}
\newcommand{\TRel}[2]{\mathrm{TRel}(#1,#2)}

\newcommand{\arity}{\mathrm{arity}}

\newcommand{\vrelated}[4]{(#1,#2) \in \Vrel{#3}{#4}}
\newcommand{\crelated}[4]{(#1,#2) \in \Crel{#3}{#4}}
\newcommand{\erelated}[4]{\crelated{#1}{#2}{#3}{#4}}
\newcommand{\vrelatedsym}[4]{(#1,#2) \in \Vrelsym{#3}{#4}}
\newcommand{\crelatedsym}[4]{(#1,#2) \in \Crelsym{#3}{#4}}
\newcommand{\erelatedsym}[4]{\crelatedsym{#1}{#2}{#3}{#4}}

\newcommand{\al}{\alpha}
\newcommand{\reverse}[1]{{#1}^{\mathrm{op}}}
\newcommand{\direction}{d}

\newcommand{\inferno}[3]{\infer[(\mbox{\textsc{#1}})]{#2}{#3}}
\newcommand{\infernoe}[3]{\infer=[(\mbox{\textsc{#1}})]{#2}{#3}}

\newcommand{\betweenrules}{\vspace{-.7ex}}

\newcommand{\eto}{\rightsquigarrow}
\newcommand{\earlier}{{\triangleleft}\,}

\newcommand{\termto}{\Downarrow}
\newcommand{\termzero}{\termto^0}

\newcommand{\bp}{.3ex}

\newcommand{\rul}[1]{\mbox{\textsc{#1}}}

%% file: defs2-stepless.tex
\newcommand{\cf}{cf. }
\newcommand{\ie}{\emph{i.e.,} }
\newcommand{\eg}{\emph{e.g.,} }
\newcommand{\etal}{\emph{et~al.}}

\newcommand{\fighead}{}
\newcommand{\figfoot}{}
\newcommand{\zilch}{\mbox{}}
\newenvironment{myfig}{\fighead}{\figfoot}
\newcommand{\mycaption}[1]{\caption{#1}}

%% file: abstract.tex
Appel and McAllester's ``step-indexed'' logical relations have proven
to be a simple and effective technique for reasoning about programs in
languages with semantically interesting types, such as general
recursive types and general reference types.  However, proofs using
step-indexed models typically involve tedious, error-prone, and
proof-obscuring step-index arithmetic, so it is important to develop
clean, high-level, equational proof principles that avoid mention of
step indices.

In this paper, we show how to reason about binary step-indexed logical
relations in an abstract and elegant way.  Specifically, we define a
logic LSLR, which is inspired by Plotkin and Abadi's logic for
parametricity, but also supports recursively defined relations by
means of the modal ``later'' operator from Appel, Melli\`es, Richards,
and Vouillon's ``very modal model'' paper.  We encode in LSLR a
logical relation for reasoning relationally about programs in
call-by-value System F extended with general recursive types.  Using
this logical relation, we derive a set of useful rules with which we
can prove contextual equivalence and approximation results without
counting steps.

%%% Local Variables: 
%%% mode: latex
%%% TeX-master: "main"
%%% End: 

%% file: intro.tex
\presec
\section{Introduction}
\label{sec:intro}
\postsec Appel and McAllester~\cite{appel-mcallester-2001} invented
the \emph{step-indexed model} in order to express ``semantic'' proofs
of type safety for use in foundational proof-carrying code.  The basic
idea is to characterize type inhabitation as a predicate indexed by
the number of steps of computation left before ``the clock'' runs out.
If a term $e$ belongs to a type $\tau$ for any number of steps (\ie
for an arbitrarily wound-up clock), then it is truly semantically an
inhabitant of $\tau$.

The step-indexed characterization of type inhabitation has the benefit
that it can be defined inductively on the step index $k$.  This is
especially useful when modeling semantically troublesome features like
recursive and mutable reference types, whose inhabitants would be
otherwise difficult to define inductively on the type structure.
Moreover, the step-indexed model's
reliance on very simple mathematical constructions makes it
particularly convenient for use in \emph{foundational} type-theoretic
proofs, in which all mathematical machinery must be mechanized.

In subsequent work, Ahmed and coworkers have shown that the
step-indexed model can also be used for \emph{relational} reasoning
about programs in languages with semantically interesting types, such
as general recursive types and general reference types~\cite{ahmed-2006,acar-ahmed-blume-2008,adr-popl09,neis+:icfp09}.

However, a continual annoyance in working with step-indexed logical
relations, as well as a stumbling block to their general acceptance,
is the tedious, error-prone, and proof-obscuring reasoning about step
indices that seems superficially to be an essential element of the
method.  To give a firsthand example: the first two authors (together
with Andreas Rossberg) recently developed a step-indexed technique for
proving representation independence of ``generative'' ADTs, \ie ADTs
that employ, in an interdependent fashion, both local state and
existential type abstraction~\cite{adr-popl09}.  While the technique
proved useful on a variety of examples, we found that our proofs using
it tended to be cluttered with step-index arithmetic, to the point
that their main substance was obscured.  Thus, it seems clear that
widespread acceptance of step-indexed logical relations will hinge on
the development of abstract proof principles for reasoning about them.

The key difficulty in developing such abstract proof principles is
that, in order to reason about things being \emph{infinitely}
logically related, \ie belonging to a step-indexed logical relation at
\emph{all} step levels---which is what one ultimately cares
about---one must reason about their
presence in the logical relation at any \emph{particular} step index,
and this forces one into finite, step-specific reasoning.  

To see a concrete example of this, consider Ahmed's step-indexed
logical relation for proving equivalence of programs written in
an extension of System F with recursive types~\cite{ahmed-2006}. One
might expect to have a step-free proof principle for establishing that
two function values are infinitely logically related, along the lines
of: $\eunf{x_1}{e_1}$ and $\eunf{x_2}{e_2}$ are infinitely logically
related at the type $\tfun{\sigma}{\tau}$ iff, whenever $v_1$ and
$v_2$ are infinitely related at $\sigma$, it is the case that
$e_1[v_1/x_1]$ and $e_2[v_2/x_2]$ are infinitely related at $\tau$.
Instead, in Ahmed's model we have that $\eunf{x_1}{e_1}$ and
$\eunf{x_2}{e_2}$ are infinitely related at $\tfun{\sigma}{\tau}$ iff
for all $n\geq 0$, whenever $v_1$ and $v_2$ are related at $\sigma$
for $n$ steps, $e_1[v_1/x_1]$ and $e_2[v_2/x_2]$ are related at $\tau$
for $n$ steps. That is, the latter is a \emph{stronger} property---if
$\eunf{x_1}{e_1}$ and $\eunf{x_2}{e_2}$ map $n$-related arguments to
$n$-related results (for any $n$), then they also map
infinitely-related arguments to infinitely-related results, but the
converse is not necessarily true. Thus, in proving infinite properties
of the step-indexed model, it seems necessary to reason about an
arbitrary finite index $n$.

In this paper, we show how to alleviate this problem by reasoning
inside a logic we call \thelogic{}.  Our approach involves a novel
synthesis of ideas from two well-known pieces of prior work: (1)
Plotkin and Abadi's logic for relational reasoning about parametric
polymorphism (hereafter, PAL)~\cite{plotkin-abadi-93}, and (2) Appel,
Melli\`es, Richards, and Vouillon's ``very modal model'' paper
(hereafter, AMRV)~\cite{appel:vmm}.

PAL is a second-order intuitionistic logic extended with axioms for
equational reasoning about relational parametricity in pure
System~F.~~Plotkin and Abadi show how to define a logical relation
interpretation of System~F types in terms of the basic constructs of
their logic.  Second-order quantification over abstract relation
variables is important in defining the relational interpretation of
polymorphic types.

In this paper, we adapt the basic apparatus of PAL toward a new
purpose: reasoning operationally about contextual equivalence and
approximation in a call-by-value language $\fmu$ with recursive and
polymorphic types.  We will show how to encode in our logic
\thelogic{} a logical relation that is sound and complete with respect
to contextual approximation, based on a step-indexed relation
previously published by Ahmed~\cite{ahmed-2006}.  Compared with
Ahmed's relation, ours is more abstract: proofs using it do not
require any step-index arithmetic.  Furthermore, whereas Ahmed's
relation is fundamentally asymmetric, our logic enables the
derivation of both equational and inequational reasoning principles.

In order to adapt PAL in this way, we need in particular the ability
to (1) reason about call-by-value and (2) logically interpret
recursive types of $\fmu$.  To address (1), we employ atomic
predicates (and first-order axioms) related to CBV reduction instead
of PAL's equational predicates and axioms.
This approach is similar to earlier logics of
partial terms for call-by-value calculi with simple~\cite{Plotkin:85} and
recursive (but not universal) types~\cite{abadi:fiore:lics96}.

For handling recursive types, it suffices to have some way of defining
recursive relations $\mu r. R$ in the logic.  This can be done when
$R$ is suitably ``contractive'' in $r$; to express contractiveness, we
borrow the ``later'' $\lift P$ operator from AMRV, which they in turn
borrowed from G\"odel-L\"ob logic~\cite{nakano-2000}.  Hence,
\thelogic{}\ is in fact not only a second-order logic (like PAL) but a
modal one, and the truth value of a proposition is the set of worlds
(think: step levels) at which it holds.
The key reasoning principle concerning the later operator is the L\"ob rule,
which states that $(\lift P \limp P) \limp P$.  This can be viewed as a
principle of induction on step levels, but we shall see that, when it is
employed in connection with logical relations, it also has a
coinductive flavor reminiscent of the reasoning principles used in
bisimulation methods like Sumii and Pierce's~\cite{sumii-pierce-jacm}.

\prepara
\subsection{Overview}
In Section~\ref{sec:lang}, we present our language under
consideration, $\fmu$.  

In Section~\ref{sec:logic}, we present our
logic \thelogic{} described above.  We give a Kripke model of
\thelogic{} with worlds being natural numbers, and ``future worlds''
being smaller numbers, so that semantic truth values are
downward-closed sets of natural numbers.  We also present a set of
basic axioms that are sound with respect to this model, and which are
useful in deriving more complex rules later in the paper.

In Section~\ref{sec:logrel}, we define a logical relation
interpretation of $\fmu$ types directly in terms of the syntactic
relations of LSLR.  Then we derive a set of useful rules for
establishing properties about the logical relation.  Using these
rules, it is easy to show that the logical relation is sound and
complete w.r.t.\ contextual approximation.  We also show in this
section how to define a symmetric version of the logical relation,
which enables direct equational reasoning about $\fmu$ programs.

In Section~\ref{sec:examples}, we give examples of contextual
equivalence proofs that employ \emph{purely logical reasoning} using
the derivable rules from Section~\ref{sec:logrel} (in particular,
without any kind of step-index arithmetic).  

In Section~\ref{sec:discussion}, we demonstrate how our LSLR proofs
improve on previous step-indexed proofs by comparing our proof for
one of the examples from Section~\ref{sec:examples} to a proof
of that example in the style of Ahmed~\cite{ahmed-2006}.

In Section~\ref{sec:lics-comparison}, we explain how the present version of LSLR
improves on (and corrects a technical flaw in) the version we
published previously in LICS~2009~\cite{dreyer+:lics09}.

Finally, in Section~\ref{sec:related}, we discuss related work
and conclude.

%%% Local Variables: 
%%% mode: latex
%%% TeX-master: "main"
%%% End: 

%% file: lang.tex
\presec
\section{The Language $\fmu$}
\label{sec:lang}
\postsec

\newcommand{\FigLangSyntax}[1][th]{
\begin{figure}[#1]
\begin{myfig}
\begin{sdisplaymath}
%\begin{array}{l}
\begin{array}{l@{\quad}r@{\quad}c@{\quad}l}
\mbox{\textit{Types}} &
\tau & \bnfdef & \alpha \bnfalt \tunit \bnfalt \tint \bnfalt \tbool \bnfalt
\tpair{\tau_1}{\tau_2} \bnfalt 
 \tsum{\tau_1}{\tau_2} \bnfalt
\tfun{\tau_1}{\tau_2} \bnfalt 
\\ & & & 
\tall{\alpha}{\tau} \bnfalt \texist{\alpha}{\tau} \bnfalt
\trec{\alpha}{\tau}  
\\[4pt] 
\mbox{\textit{Prim Ops}} &
\op & \bnfdef & + \bnfalt - \bnfalt = \bnfalt < \bnfalt \leq \bnfalt \ldots
\\[4pt] 
\mbox{\textit{Terms}} &
e & \bnfdef & x \bnfalt \eunit \bnfalt \pm\!n \bnfalt
\op(e_1,\ldots,e_n) \bnfalt  
\\ & & & 
\etrue \bnfalt \efalse \bnfalt
 \eif{e}{e_1}{e_2} \bnfalt
 \\ & & & 
\epair{e_1}{e_2} \bnfalt \efst{e} \bnfalt \esnd{e} \bnfalt 
 \\ & & & 
 \einlty{\tau}{e} \bnfalt
 \einrty{\tau}{e} \bnfalt
 \ecase{e}{x_1}{e_1}{x_2}{e_2} 
 \bnfalt
 \\ & & & 
\efun{x}{\tau}{e} \bnfalt \eapp{e_1}{e_2} \bnfalt 
\etabs{\alpha}{e} \bnfalt \etapp{e}{\tau} \bnfalt 
% \elet{x}{e_1}{e_2} \bnfalt
\\ & & & 
\epack{\tau}{e}{\texist{\alpha}{\tau'}} \bnfalt
\eunpack{e_1}{\alpha}{x}{e_2} \bnfalt 
\\ & & & 
\efoldty{\tau}{e} \bnfalt \eunfold{e} 
\\[4pt] 
\mbox{\textit{Values}} &
v & \bnfdef & x \bnfalt \eunit \bnfalt \pm\!n \bnfalt \etrue \bnfalt
\efalse \bnfalt \epair{v_1}{v_2} \bnfalt 
 \einlty{\tau}{v} \bnfalt
 \einrty{\tau}{v} \bnfalt  
\\ & & &
\efun{x}{\tau}{e} \bnfalt
\etabs{\alpha}{e} \bnfalt 
\epack{\tau_1}{v}{\texist{\alpha}{\tau}}
\bnfalt \efoldty{\tau}{v} 
\end{array}
\end{sdisplaymath}
\caption{$\fmu$ Syntax} 
\label{fig:lang:syntax}
\end{myfig}
\end{figure}
}

\FigLangSyntax[t]

\newcommand{\AFigLangDynamicSem}[1][th]{
\begin{figure}[#1]
\begin{myfig}
%\vspace{-2ex}
\begin{sdisplaymath}
\begin{array}{l@{\quad}r@{\quad}c@{\quad}l}
\mbox{\textit{Eval. Contexts}} &
\ectxt & \bnfdef & 
\hole \bnfalt 
\op(v_1,\ldots,v_{i-1},\ectxt,e_{i+1},\ldots,e_n)
\bnfalt 
\\&&&
\eif{\ectxt}{e_1}{e_2} \bnfalt 
\epair{\ectxt}{e_2} \bnfalt \epair{v_1}{\ectxt} \bnfalt 
\efst{\ectxt} \bnfalt \esnd{\ectxt} \bnfalt 
\\ &&&
\einlty{\tau}{\ectxt} \bnfalt \einrty{\tau}{\ectxt} \bnfalt
\ecase{\ectxt}{x_1}{e_1}{x_2}{e_2} \bnfalt  
\\ &&&
\eapp{\ectxt}{e} \bnfalt \eapp{v}{\ectxt} \bnfalt \etapp{\ectxt}{\tau} 
\bnfalt  
%\\ &&&
\epack{\tau_1}{\ectxt}{\texist {\alpha}{\tau}} \bnfalt
\eunpack{\ectxt}{\alpha}{x}{e_2} \bnfalt 
\\ &&&
\efoldty{\tau}{\ectxt} \bnfalt \eunfold{\ectxt} %\bnfalt
\end{array}
\end{sdisplaymath}

%\hrule

\noindent
\hspace{0.34in}\fbox{\small$e \step e'$} \hfill
\begin{sdisplaymath}
\begin{array}{@{}r@{~}c@{~}l}
 \eif{\etrue}{e_1}{e_2} & \step &  e_1 \\
 \eif{\efalse}{e_1}{e_2} & \step &  e_2 \\
 \efst{\epair{v_1}{v_2}} & \step &  v_1 \\
 \esnd{\epair{v_1}{v_2}} & \step &  v_2 \\
 \ecase{(\einlty{\tau}{v})}{x_1}{e_1}{x_2}{e_2} & \step & 
\subst{v}{x_1}{e_1} \\ 
 \ecase{(\einrty{\tau}{v})}{x_1}{e_1}{x_2}{e_2} & \step & 
\subst{v}{x_2}{e_2} \\ 
 \eapp{(\efun{x}{\tau}{e})}{v} & \step &  \subst{v}{x}{e} \\
% \elet{x}{v}{e} & \step &  \subst{v}{x}{e} \\
 \etapp{(\etabs{\alpha}{e})}{\tau} & \step & 
\subst{\tau}{\alpha}{e} \\
 \eunpack{(\epack{\tau}{v}{\texist {\alpha}{\tau_1}})}{\alpha}{x}{e} 
& \step &  \subst{\tau}{\alpha}{\subst{v}{x}{e}} \hspace{0.8in}\\
 \eunfold{(\efoldty{\tau}{v})} & \step &  v
\end{array}
\end{sdisplaymath}
%\vspace{-0.9ex}
\begin{smathpar}
\inferrule{e ~~\step~~ e'}
{\ectxt\hw{e} ~~\step~~ \ectxt\hw{e'}}
\end{smathpar}
\caption{$\fmu$ Dynamic Semantics} 
\label{fig:apdx:dynamicsem}
\end{myfig}
\end{figure}
}

\AFigLangDynamicSem[t]

We consider $\fmu$, a call-by-value $\lambda$-calculus with
impredicative polymorphism and iso-re\-cur\-sive types.  The syntax of
$\fmu$ is shown in Figure~\ref{fig:lang:syntax}.  Sum and recursive
type injections are type-annotated to ensure unique typing, but we
will often omit the annotations when they are obvious from context.
Figure~\ref{fig:apdx:dynamicsem} shows the left-to-right call-by-value
dynamic semantics for the language, defined as a small-step relation
on terms (written $e \step e'$), which employs evaluation contexts $E$
in the standard way.  Note that the reduction relation is
deterministic.

$\fmu$ typing judgments have the form $\Gamma \ts e : \tau$, where the
context $\Gamma$ binds type variables $\alpha$, as well as
term variables~$x$:
$\Gamma~\bnfdef~\empctx \bnfalt \Gamma, \alpha \bnfalt \Gamma, x:\tau$.
The typing rules are also standard and are given in full in
Appendix~\ref{sec:apdx:lang} (Figure~\ref{fig:apdx:staticsem}).  

\subsection{Contextual Approximation and Equivalence}
\label{sec:lang-contextual}
A context $C$ is a term with a single hole $\hole$ in it.  The typing
judgment for contexts has the form $\vdash C : (\twf{\Gamma}{\tau})
\ctxarrow (\twf{\Gamma'}{\tau'})$, where $(\twf{\Gamma}{\tau})$
indicates the type of the hole.  This judgment essentially says that
if $e$ is a term such that $\judg{\Gamma}{e}{\tau}$, then
$\judg{\Gamma'}{C\hw{e}}{\tau'}$.  
Its formal definition appears in %the appendix~\cite{appendix}.
Appendix~\ref{sec:apdx:lang} (Figures~\ref{fig:apdx:contextsI}
and~\ref{fig:apdx:contextsII}). 

We define contextual approximation
(\mbox{$\judg{\Gamma}{e_1{\,\ctxaprx\,}e_2}{\tau}$}) to mean that, for
any well-typed program context $C$ with a hole of the type of $e_1$
and $e_2$, the termination of $C[e_1]$ (written $C[e_1] \Downarrow$)
implies the termination of $C[e_2]$. Contextual equivalence
($\judg{\Gamma}{e_1 \ctxeqv
  e_2}{\tau}$) is then defined as approximation in both directions.
\begin{defi}[Contextual Approximation \& Equivalence] Let
  $\judg{\Gamma}{e_1}{\tau}$ and $\judg{\Gamma}{e_2}{\tau}$.  
\predisp
\[
%\begin{small}
\begin{array}{l}
\judg{\Gamma}{e_1 \ctxaprx e_2}{\tau} ~~\defeq~~
\vspace{2pt}
\begin{stackTL}
\forall C,\tau'\!.~\,\begin{stackTL}(\ts C :
(\twf{\Gamma}{\tau}) \!\ctxarrow\! 
(\twf{\empctx}{\tau'}) \land  \termin{C\hw{e_1}}) \ \limp \termin{C\hw{e_2}} 
\end{stackTL}
\end{stackTL}\\
\judg{\Gamma}{e_1 \ctxeqv e_2}{\tau} ~~\defeq~~
\judg{\Gamma}{e_1 \ctxaprx e_2}{\tau} ~\land~
\judg{\Gamma}{e_2 \ctxaprx e_1}{\tau}
\end{array}
\postdisp
%\end{small}
\]
\end{defi}

%%% Local Variables: 
%%% mode: latex
%%% TeX-master: "main"
%%% End: 

%% file: logic.tex
\presec
\section{The Logic \thelogic{}}
\label{sec:logic}
\postsec
\thelogic{} is a second-order intuitionistic modal logic supporting a
primitive notion of term relations, as well as the ability to
define such relations recursively.

\begin{figure}[!t]
\begin{myfig}
\begin{sdisplaymath}
\begin{array}{l@{\ }r@{\quad}c@{\quad}l}
\mbox{\textit{Relation Variables}} & r & \in & \mathit{RelVar} \\
\mbox{\textit{$\fmu$ Variable Contexts}} & \vcon & \bnfdef &
  \empctx \bnfalt \vcon,\alpha \bnfalt \vcon,x \\
\mbox{\textit{$\fmu$ Variable  Substitutions}} & \gamma & \bnfdef &
  \empctx \bnfalt \gamma,\alpha\mapsto\tau \bnfalt \gamma,x\mapsto e \\
\mbox{\textit{Relation Contexts}} & \rcon & \bnfdef &
  \empctx \bnfalt \rcon,r \\
\mbox{\textit{Relation Substitutions}} & \varphi & \bnfdef &
  \empctx \bnfalt \varphi,r\mapsto R \\
\mbox{\textit{Proposition Contexts}} & \acon & \bnfdef & \empctx \bnfalt \acon,P \\
\mbox{\textit{Combined Contexts}}& \con & \bnfdef & \vcon;\rcon;\pcon \\
\mbox{\textit{Atomic Relations}} & A,B & \bnfdef & e_1 = e_2 \bnfalt \cdots \\
\mbox{\textit{Relations}} & P,Q,R,S & \bnfdef & r \bnfalt 
  A \bnfalt \top \bnfalt \bot \bnfalt
  P \land Q \bnfalt P \lor Q \bnfalt P \limp Q \bnfalt  \\&&&
  \forall \vcon.P \bnfalt \exists \vcon.P \bnfalt 
  \forall \rcon.P \bnfalt \exists \rcon.P \bnfalt \\&&&
  \seq{x}.P \bnfalt \seq{e} \in R \bnfalt \mu r.R \bnfalt \lift P
  \\
\end{array}
\end{sdisplaymath}
\mycaption{Syntax of Core \thelogic{}}
\label{fig:logic-syntax}
\end{myfig}
\end{figure}

%\vspace{-2ex}
\subsection{Syntax}

The core syntax of \thelogic{} is given in Figure~\ref{fig:logic-syntax}.

$\fmu$ variable contexts $\vcon$ are similar to $\fmu$ contexts
$\Gamma$, except that they omit type annotations on term
variables.  Instead, well-typedness of variables is modeled through
explicit typing hypotheses in the proposition context $\pcon$ (see
below).  $\fmu$ variable substitutions $\gamma$ map variables bound in
$\fmu$ variable contexts to objects of the appropriate syntactic
class.

As a matter of notation, we will use $y$ and $t$ as term variables in
addition to $x$.  Often, we write $x$ or $y$ to denote values, whereas
$t$ stands for arbitrary terms.  (This is merely a mnemonic, however.
The fact that $x$ or $y$ is a value will always be guaranteed by some
separate, explicit assumption.)

Relation contexts $\rcon$ bind relation variables $r$, which stand for
relations of arbitrary arity between $\fmu$ terms.  For ease of
notation, we assume that relation variables $r$ come equipped
implicitly with a particular arity (namely, $\arity(r)$).  Relation
substitutions $\varphi$ map relation variables to \emph{relations} $R$
of the appropriate arity, which we describe below.  

Proposition contexts $\pcon$ are sets of \emph{propositions},
which are just nullary relations that we typically denote using $P$
and $Q$.  (Note: We treat all three kinds of contexts as unordered
sets, and use comma to denote disjoint union of such sets.)

We write $\con$ to denote a combined context $\vcon;\rcon;\pcon$.
Correspondingly, we also define $\con,\vcon'$ to mean
$\vcon,\vcon';\rcon;\pcon$ (and similarly for $\con,\rcon'$ and
$\con,\pcon'$).

Relations $R$ (of which propositions $P$ are a subset) fall into
several categories: \emph{variable} relations ($r$), \emph{atomic}
relations ($A$), \emph{first-order} propositions ($\top$, $\bot$, $P
\land Q$, $P \lor Q$, $P \limp Q$, $\forall\vcon.P$,
$\exists\vcon.P$), \emph{second-order} propositions ($\forall\rcon.P$,
$\exists\rcon.P$), relation introduction and elimination ($\seq{x}.P$,
$\seq{e}\in R$), recursive relations ($\mu r. R$), and the
\emph{later} modality ($\lift P$) borrowed from AMRV~\cite{appel:vmm}.

Atomic propositions $A$ and the axioms concerning them are essentially
orthogonal to the other components of the logic.  We have listed in
Figure~\ref{fig:logic-syntax} one particularly central atomic
proposition, $e_1 = e_2$, which says that $e_1$ and $e_2$ are
syntactically equal modulo renaming of bound
variables.  % (up to $\alpha$-equivalence).
In Section~\ref{sec:atomic}, we will introduce several other atomic
propositions related to the reduction semantics of $\fmu$.  The only
common requirement we impose on all of these atomic propositions is
that they are first-order, in the sense that they only depend on type
and term variables, not relation variables.

The first-order connectives are self-explanatory.  The second-order
ones provide the ability to abstract over a relation, which is
critical in defining logical relations for polymorphic and existential
types.  As for the relational introduction and elimination forms:
$\seq{x}.P$, which we sometimes write as $(\seq{x}).P$, introduces the
term relation that one would write in set notation as $\{(\seq{x})
\mid P\}$, and $\seq{e} \in R$ says that the tuple of terms
$(\seq{e})$ belong to the relation $R$.  In general, we use the
overbar notation to denote a possibly nullary tuple of objects.

A recursive relation $\mu r.R$ denotes the relation $R$ that may refer
to itself recursively via the variable $r$.  In order to ensure that
such relations are well-founded, we require that $R$ be
\emph{contractive} in $r$, a notion that we make precise (following
AMRV) using the modal $\later$ operator.  Specifically, we define $R$
to be contractive in $r$ if $r$ may only appear in $R$ underneath the
$\later$ operator (\ie inside propositions of the form $\later P$).
Intuitively (and formally), $\later P$ means that $P$ is true in all
\emph{strictly} future worlds of the current one.  As a result, the
meaning of $\mu r.R$ only depends recursively on its own meaning in
strictly future worlds.  Thus, assuming that the ``strictly future
world'' ordering is well-founded, we can define the meaning of $\mu
r.R$ by induction on strictly future worlds.

\subsection{A ``Step-Indexed'' Model of LSLR}
\label{sec:model}

\begin{figure}[!t]
\begin{myfig}
\small
\noindent
\hspace{1in} If $n=0$, then: \hfill \mbox{}
\vspace{-2ex}
\[
\begin{array}{r@{\quad\defeq\quad}l}
\interp{R}{n}{\delta}\seq{e} & \top \hspace{11ex} \\
\interp{\acon}{n}{\delta} & \top \\
\end{array}
\]

\hspace{1in} If $n>0$, then: \hfill \mbox{}
\vspace{-2ex}
\[
\begin{array}{r@{\quad\defeq\quad}l}
\interp{r}{n}{\delta}\seq{e} & \delta rn\seq{e} \\
\hspace{11.8ex}\interp{A}{n}{\delta}\seq{e} & \mathcal{I}(A)\seq{e} \\
\interp{\top}{n}{\delta} & \top \\
\interp{\bot}{n}{\delta} & \bot \\
\interp{P \land Q}{n}{\delta} & \interp{P}{n}{\delta} \land \interp{Q}{n}{\delta} \\
\interp{P \lor Q}{n}{\delta} & \interp{P}{n}{\delta} \lor \interp{Q}{n}{\delta} \\
\interp{P \limp Q}{n}{\delta} & \forall k \leq n.\ \interp{P}{k}{\delta} \limp \interp{Q}{k}{\delta} \\
\interp{\forall \vcon. P}{n}{\delta} & \forall \gamma \in \sembrace{\vcon}.\ \interp{\gamma P}{n}{\delta} \\
\interp{\exists \vcon. P}{n}{\delta} & \exists \gamma \in \sembrace{\vcon}.\ \interp{\gamma P}{n}{\delta} \\
\interp{\forall \rcon. P}{n}{\delta} & \forall \delta' \in \sembrace{\rcon}.\ \interp{P}{n}{(\delta,\delta')} \\
\interp{\exists \rcon. P}{n}{\delta} & \exists \delta'\in \sembrace{\rcon}.\ \interp{P}{n}{(\delta,\delta')} \\
\interp{\seq{x}.P}{n}{\delta}\seq{e} &
  \interp{P[\seq{e}/\seq{x}]}{n}{\delta} \\
\interp{\seq{e} \in R}{n}{\delta} & \interp{R}{n}{\delta}\seq{e} \\
\interp{\mu r. R}{n}{\delta}\seq{e} &
  \interp{R[\mu r. R/r]}{n}{\delta}\seq{e} \\
\interp{\lift P}{n}{\delta} & \interp{P}{(n-1)}{\delta} \\[1.5ex]
\interp{\pcon}{n}{\delta} & \forall P \in \pcon.\ \interp{P}{n}{\delta} \\
\end{array}
\]
\mycaption{Kripke ``Step-Indexed'' Model of \thelogic{}}
\label{fig:model}
\end{myfig}
\end{figure}

Figure~\ref{fig:model} defines a Kripke model for \thelogic{}, where
the worlds are natural numbers and $m$ is a strictly future world of
$n$ if $m<n$.  The model enjoys \emph{monotonicity}, meaning that if a
proposition is true in world $n$, it is true in all strictly future
worlds as well.  Thus, the set of semantic truth values is the
complete Heyting algebra $\mathcal{P}^\downarrow(\mathbb{N})$ of
downward-closed subsets of $\mathbb{N}$, ordered by inclusion (or,
isomorphically, the complete Heyting algebra $\vec{\omega}$ of
vertical natural numbers with infinity).  

We interpret relations and proposition contexts under some semantic
interpretation $\delta$, which maps their free relation variables to
\emph{semantic} (\ie world-indexed, monotone) relations of the
appropriate arity.  We write $\interp{R}{n}{\delta}\seq{e}$ (resp.\
$\interp{\acon}{n}{\delta}$) to mean that, under interpretation
$\delta$, $\seq{e} \in R$ (resp.\ $\acon$) is true in world $n$.  The
interpretations refer to $\sembrace{\vcon}$ and $\sembrace{\rcon}$.
The semantic interpretation of a variable context, $\sembrace{\vcon}$,
is the set of closing variable substitutions $\gamma$ whose domains
equal $\vcon$.  The semantic interpretation of a relation context,
$\sembrace{\rcon}$, is the set of semantic relation substitutions
$\delta$ whose domains equal $\rcon$.

The interpretations in Figure~\ref{fig:model} are defined by a double
induction, first on the world $n$ (in world $0$, everything is true),
and second on the ``size'' of the relation being interpreted.  The
\emph{size} of a relation is defined to equal the number of
logical/relational connectives in it, \emph{ignoring} all connectives
appearing inside a proposition of the form $\later P$ (\ie $\later P$
has constant size, no matter what $P$ is).  This size metric makes it
possible to interpret a recursive relation $\mu r.R$ directly in terms
of its expansion $R[\mu r.R/r]$.  Assuming the relation is
well-formed, this interpretation is well-defined because the expansion
has a smaller size.  (Specifically, since $R$ is contractive in $r$,
we know that $r$ may only appear inside constant-size propositions in
$R$, so the size of $R[\mu r.R/r]$ equals the size of $R$, which is
smaller than the size of $\mu r.R$.)

Since $\later P$ may have smaller size than $P$, it is critical
that the interpretation of $\later P$ in world $n$ be defined in terms
of the interpretation of $P$ in strictly future worlds (\ie worlds
strictly less than $n$).  Fortunately, this is no problem since, as
explained above, $\later P$ means precisely that $P$ is true in all
strictly future worlds.  Thanks to the built-in monotonicity
restriction, it suffices to say that $\later P$ is true in world $n$
iff $P$ is true in world $n-1$.

Otherwise, the interpretation is mostly standard.  One point of note
is the interpretation of implication $P \limp Q$, which quantifies
over all future worlds in order to ensure monotonicity.
Another is the interpretation of atomic relations $A$.  We assume an
interpretation function $\mathcal{I}$, which maps closed atomic
relations $A$ to \emph{absolute} (\ie world-independent) relations.
As one instance, we define $\mathcal{I}(e_1 = e_2)$ to be true
($\top$) iff $e_1$ is $\alpha$-equivalent to $e_2$.

Using this model, we can define our main logical judgment,
$\vcon;\rcon;\acon \ts P$.  Assuming that $\acon$ and $P$ are
well-formed in $\vcon;\rcon$ (see Appendix~\ref{sec:apdx:logic} for
the definition of proposition/relation well-formedness), the judgment
is interpreted as follows:
\[
\vcon;\rcon;\acon \ts P \ \ \defeq\ \ \forall n \geq 0.\ \forall
\gamma\in\sembrace{\vcon}.\ \forall \delta\in\sembrace{\rcon}.\
\interp{\gamma \acon}{n}{\delta} ~\limp~ \interp{\gamma P}{n}{\delta}
\]
Note that we interpret the judgment directly as a statement in the
model, rather than inductively defining it via a set of inference
rules.  This allows us to prove new inference rules sound whenever
needed.  In the next section, however, we will establish a core set of
sound inference rules that will enable us to reason about the judgment
(in most cases) without having to appeal directly to the model.

The judgment asserts that under any closing substitution $\gamma$ for
$\vcon$ and any semantic interpretation $\delta$ for $\rcon$, and in
any world $n$, the hypotheses $\pcon$ imply the conclusion $P$.  The
key here is that, while $n$ is universally quantified and thus not
explicitly mentioned in the logical judgment, the hypotheses $\pcon$
and the conclusion $P$ are both interpreted in the \emph{same} world
(\ie step level) $n$.  This is what allows us to prove something like
``$f_1$ and $f_2$ map $n$-related arguments to $n$-related results''
(as discussed in the introduction) without having to talk about a
specific step level $n$.

Finally, it is worth noting that, while the Kripke model we have
defined here may be viewed as a ``step-indexed'' model, nothing in the
model mentions steps of computation.  We happen to be using natural
numbers as our worlds, but there is no computational meaning attached
to them at this point.  The connection between worlds and (certain)
steps of computation will be made later on, when we define the logical
relation for $\fmu$ in Section~\ref{sec:logrel}.

\begin{figure}[!t]
\begin{myfig}
\small

\[
\inferno{mono}{
\con \ts \later P
}{
\con \ts P
}
\qquad
\inferno{l\"ob}{
\con \ts P
}{
\con, \later P \ts P
}
\]
\betweenrules
\[
\infernoe{$\later{\land}$}{
\con \ts \later P \land \later Q
}{
\con \ts \later (P \land Q)
}
\qquad
\infernoe{$\later{\lor}$}{
\con \ts \later P \lor \later Q
}{
\con \ts \later (P \lor Q)
}
\qquad
\infernoe{$\later{\limp}$}{
\con \ts \later P \limp \later Q
}{
\con \ts \later (P \limp Q)
}
\]
\betweenrules
\[
\infernoe{$\later{\forall}1$}{
\con \ts \forall \vcon. \later P
}{
\con \ts \later \forall \vcon. P
}
\qquad
\infernoe{$\later{\exists}1$}{
\con \ts \exists \vcon. \later P
}{
\con \ts \later \exists \vcon. P
}
\qquad
\infernoe{$\later{\forall}2$}{
\con \ts \forall \rcon. \later P
}{
\con \ts \later \forall \rcon. P
}
\qquad
\infernoe{$\later{\exists}2$}{
\con \ts \exists \rcon. \later P
}{
\con \ts \later \exists \rcon. P
}
\]
\betweenrules
\[
\inferno{replace1}{
\con \ts P[e_2/x]
}{
\con \ts e_1 = e_2 \quad
\con \ts P[e_1/x]
}
\qquad
\inferno{replace2}{
\con \ts P[R_2/r]
}{
\con \ts R_1 \equiv R_2 \quad
\con \ts P[R_1/r]
}
\]
\betweenrules
\[
\infernoe{elem}{
      \con \ts P[\seq{e}/\seq{x}]
    }{
      \con \ts \seq{e} \in \seq{x}.P
    }
    \qquad
\infernoe{elem-$\mu$}{
      \con \ts \seq{e} \in R[\mu r.R/r]
    }{
      \con \ts \seq{e} \in \mu r.R
    }
\]
\mycaption{Core Inference Rules of \thelogic{}}
\label{fig:logic-rules}
\end{myfig}
\end{figure}

\subsection{Core Inference Rules}
\label{sec:core-rules}

We now present the core inference rules of LSLR, all of which are easy
to prove sound directly in the model.  The most interesting ones are
shown in Figure~\ref{fig:logic-rules}; the remainder, all of which are
standard rules for second-order intuitionistic logic, appear in
Appendix~\ref{sec:apdx:logic}.

Rule~\rul{mono} is the axiom of monotonicity, stating that
propositions that are true now (in the current world) are also true
later (in future worlds).  The \rul{l\"ob} rule, adapted from AMRV,
provides a clean induction principle over future worlds.  If under the
assumption that $A$ is true later (in all strictly future worlds) we
can prove that it is true in the current world, then by induction $A$
is true in the current world.  The induction argument requires no base
case because all propositions are assumed true in the final world (\ie
world $0$).

The remainder of the rules concerning the later operator state that
the later operator distributes over all propositional connectives.
Not all these distributivity laws are valid in classical G\"odel-L\"ob
logic or AMRV, but they hold here due to our axiom of monotonicity.
For example, we give here the proof of Rule~\rul{$\later{\limp}$}:

\begin{prop}
Rule~\rul{$\later{\limp}$} is admissible.
\end{prop}

\proof First, the forwards direction.  Suppose $\interp{\later(P \limp
  Q)}{n}{\delta}$ and $\interp{\later{P}}{n}{\delta}$.  We want to
show $\interp{\later{Q}}{n}{\delta}$.  If $n=0$, the proof is trivial,
so assume $n > 0$.  By the interpretation of $\later$, we know
$\interp{P \limp Q}{(n-1)}{\delta}$ and $\interp{P}{(n-1)}{\delta}$.
Thus, by the interpretation of $\limp$, we know
$\interp{Q}{(n-1)}{\delta}$, which is equivalent to our goal.

Next, the backwards direction.  Suppose
$\interp{\later{P}\limp\later{Q}}{n}{\delta}$; we want to show
$\interp{\later(P \limp Q)}{n}{\delta}$.  If $n=0$, the proof is
trivial, so assume $n > 0$.  Our goal is equivalent to $\interp{P
  \limp Q}{(n-1)}{\delta}$, so suppose $k \leq n-1$ and
$\interp{P}{k}{\delta}$, and we will prove $\interp{Q}{k}{\delta}$.
By the interpretation of $\later$, we know $\interp{\later
  P}{(k+1)}{\delta}$.  Since $k+1 \leq n$, by the interpretation of
$\limp$ we obtain $\interp{\later Q}{(k+1)}{\delta}$, which is
equivalent to $\interp{Q}{k}{\delta}$, our desired goal.  

Note that the backwards direction relies critically on monotonicity.
In the absence of monotonicity, the premise
$\interp{\later{P}\limp\later{Q}}{n}{\delta}$ is only applicable if
$\interp{P}{k}{\delta}$ for \emph{all} $k<n$, but in the proof we only
assume $\interp{P}{k}{\delta}$ for \emph{some} $k<n$.  
\qed

The replacement axioms (\rul{replace1} and \rul{replace2}) say that we
can substitute equals for equals inside a proposition without
affecting its meaning.  For terms, equality is just syntactic
equality.  For relations, equivalence is definable as
\[
R_1 \equiv R_2 \ \ \defeq\ \ \forall \seq{x}.\ (\seq{x} \in R_1 \limp \seq{x} \in R_2) ~\land~
(\seq{x} \in R_2 \limp \seq{x} \in R_1)
\]

The last two rules concern inhabitation of relations.  The key
interesting point here is that recursive relations are equivalent to
their expansions.

Lastly, when we introduce atomic propositions in the next section
related to $\fmu$ reduction, we will want to also import into LSLR
various first-order theorems about those propositions, \eg
preservation, progress, canonical forms, etc.  Fortunately, this can
be done easily, without requiring any stepwise reasoning.

Formally, assuming $P$ is a first-order proposition (\ie it does not
involve relation variables, recursive relations, second-order
quantification, or the $\later$ operator), then it is easy to show
that $P$ is true in all worlds $n$ iff it is true in world $1$ (the
``latest'' nontrivial world).  Consequently, the following rule is sound:
\[
\infer{
\vcon;\rcon;\pcon \ts P
}{
\forall \gamma\in\sembrace{\vcon}.\ \forall \delta\in\sembrace{\rcon}.\ \sembrace{\gamma\acon}\delta(1) \limp \sembrace{\gamma P}\delta(1)
}
\postdisp
\]
Thus, in particular, if $P$ is closed:
\[
\infer{
\ts P
}{
\sembrace{P}1
}
\]
For first-order $P$, the interpretation of $\sembrace{P}1$ in our
model is tantamount to the standard step-free interpretation of $P$ in
first-order logic.

In other words, our goal here is not to use \thelogic{} to formalize
\emph{entire} proofs, just the parts of the proofs that involve
interesting relational reasoning.  We are happy to make use of
first-order syntactic properties proved by other means in the
meta-logic.

%%% Local Variables: 
%%% mode: latex
%%% TeX-master: "main"
%%% End: 

%% file: logrel.tex
\presec
\section{A Syntactic Logical Relation for $\fmu$}
\label{sec:logrel}
\postsec

In this section, we show how to define a logical relation for $\fmu$
that coincides with contextual approximation, as well as a symmetric
version thereof that coincides with contextual equivalence.  The
relation is defined \emph{syntactically} within the logic \thelogic{},
using a particular set of atomic propositions concerning the $\fmu$
reduction semantics, as we explain below.

\subsection{Roadmap and Preliminaries}

Eventually, we are going to define a logical relation on open terms,
which we denote $\Gamma \ts e_1 \logaprx e_2 : \tau$, and prove that
it is sound and complete w.r.t.\ contextual approximation, $\Gamma \ts
e_1 \ctxaprx e_2 : \tau$, as defined in Section~\ref{sec:lang}.  In
order to prove this, we will follow Pitts~\cite{pitts:attapl} in
employing an intermediate form of approximation, often referred to as
\emph{ciu approximation}.

Ciu approximation, due to Mason and Talcott~\cite{mason-talcott-91}, is a
superficially coarser version of contextual approximation in which (1)
attention is restricted to evaluation contexts $E$ instead of
arbitrary program contexts, and (2) the ``closing'' of open terms is
handled by an explicit substitution $\gamma$ instead of relying on
$\lambda$-abstractions in a closing context $C$.  We say that ciu
approximation is only \emph{superficially} coarser because ultimately
we will prove that it too coincides with contextual approximation.  In
the meantime, ciu approximation turns out to be an easier notion of
approximation to work with.

First, a bit of notation: we will write $\ts \gamma : \Gamma$ to mean
that (1) $\dom(\gamma) = \dom(\Gamma)$, (2) $\forall\alpha\in\Gamma.\
\fv(\gamma\alpha) = \emptyset$, and (3) $\forall x:\tau\in\Gamma.\
\exists v.\ \gamma x = v ~\land~ \ts v : \gamma\tau$.  We will also
write $\ts E : \tau \eto \tau'$ to mean $\ts E : (\twf{\empctx}{\tau})
\eto (\twf{\empctx}{\tau'})$, thus defining the typing of evaluation
contexts in terms of the typing judgment for general contexts $C$
(introduced in Section~\ref{sec:lang-contextual}).

\begin{defi}[Ciu Approximation for Closed Terms] Let
  $\judg{\empctx}{e_1}{\tau}$ and $\judg{\empctx}{e_2}{\tau}$.  
\predisp
\[
%\begin{small}
\begin{array}{l}
\ts e_1 \ciuaprx e_2 : \tau ~~\defeq~~ 
\vspace{2pt}
\begin{stackTL}
\forall E,\tau'\!.~\,\begin{stackTL}(\ts E :
\tau \eto \tau' \land \termin{E\hw{e_1}}) \ \limp \termin{E\hw{e_2}} 
\end{stackTL}
\end{stackTL}\\
\end{array}
\postdisp
%\end{small}
\]
\end{defi}

\begin{defi}[Ciu Approximation for Open Terms] Let
  $\judg{\Gamma}{e_1}{\tau}$ and $\judg{\Gamma}{e_2}{\tau}$.  
\predisp
\[
%\begin{small}
\begin{array}{l}
\judg{\Gamma}{e_1 \ciuaprx e_2}{\tau} ~~\defeq~~ 
\vspace{2pt}
\begin{stackTL}
\forall \gamma.\ \ts \gamma : \Gamma ~\limp~ {\ts}~ \gamma e_1 \ciuaprx \gamma e_2 : \gamma\tau
\end{stackTL}\\
\end{array}
\postdisp
%\end{small}
\]
\end{defi}

\begin{defi}[Ciu Equivalence] 
  Let $\judg{\Gamma}{e_1}{\tau}$ and $\judg{\Gamma}{e_2}{\tau}$.
\[
\judg{\Gamma}{e_1 \ciueqv e_2}{\tau} ~~\defeq~~ \judg{\Gamma}{e_1
    \ciuaprx e_2}{\tau} ~\land~ \judg{\Gamma}{e_2 \ciuaprx e_1}{\tau}
\]
\end{defi}

\noindent
One of the main reasons to use ciu approximation instead of contextual
approximation is that it is immediately obvious that the $\fmu$
reduction relation is contained in ciu equivalence (part (3) of
the following proposition).

\begin{prop}[Useful Properties of Ciu Approximation]
\label{prop:ciu-props} \hfill
\begin{enumerate}[\em(1)]
\item
If $\Gamma \ts e : \tau$, then $\Gamma \ts e \ciuaprx e : \tau$.
\item
If $\Gamma \ts e_1 \ciuaprx e_2 : \tau$ and $\Gamma \ts e_2 \ciuaprx e_3 : \tau$,
then $\Gamma \ts e_1 \ciuaprx e_3 : \tau$.
\item
If $\Gamma \ts e_1 : \tau$ and $e_1 \stepmany e_2$, then $\Gamma \ts e_1 \ciueqv e_2 : \tau$.
\item
If $\ts e_1 \ciuaprx e_2 : \tau$ and $\ts E : \tau \eto \tau'$, then
$\ts E[e_1] \ciuaprx E[e_2] : \tau'$.
\end{enumerate}
\end{prop}

\noindent
Again following Pitts~\cite{pitts:attapl}, we will show that contextual, ciu,
and logical approximation all coincide by showing that
$\ctxaprx ~\subseteq~ \ciuaprx ~\subseteq~ \logaprx ~\subseteq~ \ctxaprx$.
The first link of that chain is easy.

\begin{thm}[Contextual Approximation $\limp$ Ciu Approximation]
\label{thm:ctx-ciu} \hfill \\
If $\Gamma \ts e_1 \ctxaprx e_2 : \tau$, then $\Gamma \ts e_1 \ciuaprx e_2 : \tau$.
\end{thm}

\proof Suppose $\ts \gamma : \Gamma$, $\ts E : \gamma\tau \eto \tau'$,
and $E[\gamma e_1] \termto$.  We want to show $E[\gamma e_2] \termto$.
Say that $\Gamma =
\alpha_1,\ldots,\alpha_m,x_1\,{:}\,\tau_1,\ldots,x_n\,{:}\,\tau_n$ and
that $\gamma\alpha_i = \sigma_i$ and $\gamma x_i = v_i$ for some
$\sigma_i$'s and $v_i$'s.  Then, let $C =
(\Lambda\alpha_1.\cdots\Lambda\alpha_m.\lambda
x_1\,{:}\,\tau_1.\cdots\lambda
x_n\,{:}\,\tau_n.\,\hole)\,\sigma_1\cdots\sigma_mv_1\cdots v_n$.  It
is easy to show that $\ts C : (\Gamma \ts \tau) \eto (\empctx \ts
\gamma\tau)$, and thus that $\ts E[C] : (\Gamma \ts \tau) \eto
(\empctx \ts \tau')$.  It is also easy to show that
$E[C[e_i]] \stepmany E[\gamma e_i]$, and thus that $E[C[e_i]] \termto$
iff $E[\gamma e_i] \termto$.  So the goal is reduced to showing that
$E[C[e_1]] \termto$ implies $E[C[e_2]] \termto$, which follows from
$\Gamma \ts e_1 \ctxaprx e_2 : \tau$.
\qed

\vspace{-2ex}
\subsection{Atomic Relations}
\label{sec:atomic}

In order to define our logical relation, we introduce the following
new atomic relations:
\[
A\ ::=\ \cdots \bnfalt \OVal \bnfalt e : \tau \bnfalt C : \tau \eto \tau' \bnfalt e_1 \evalsto e_2 \bnfalt e_1 \stepzero e_2 \bnfalt e_1 \stepone e_2 \bnfalt e_1 \ciuleq e_2
\]
Except for the first, which is a unary relation, the rest are all
nullary (\ie propositions).  The interpretations of these
propositions, $\atominterp{A}$, are as follows:
\begin{enumerate}[$\bullet$]
\item
$\atominterp{\OVal}(e)$ $\defeq$ $\exists v.\ e = v$.
\item
$\atominterp{e : \tau}$ $\defeq$ $\ts e : \tau$.
\item
$\atominterp{C : \tau \eto \tau'}$ $\defeq$ $\exists E.\ C = E ~\land~ {\ts}~ E : \tau \eto \tau'$.
\item
$\atominterp{e_1 \evalsto\! e_2}$ $\defeq$ $e_1 \stepmany e_2$.
\item
$\atominterp{e_1 \stepzero e_2}$ $\defeq$ $e_1 \stepmany e_2$
and \emph{none} of the reductions in the reduction sequence is an \cd{unroll}-\cd{roll}
reduction.\\[-2ex]
\item $\atominterp{e_1 \stepone e_2}$ $\defeq$ $e_1 \stepmany e_2$ and
  \emph{exactly one} of the reductions in the reduction sequence is an
  \cd{unroll}-\cd{roll} reduction.\\[-2ex]
\item
$\atominterp{e_1 \ciuleq e_2}$ $\defeq$ $\exists \tau.\ \ts e_1 \ciuaprx e_2 : \tau$.
\end{enumerate}

The motivation for using this particular set of atomic propositions
will become clear shortly.  One point of note is that the $e_1 \ciuleq
e_2$ proposition lacks a type; this is simply for brevity, since
$\fmu$ enjoys unique typing.  Another is that, although the
proposition $C : \tau \eto \tau'$ permits an arbitrary context $C$,
the proposition only holds when $C$ takes the form of an evaluation
context, and we will only use it when $C$ is an evaluation context.
The reason that we do not syntactically write $E$ here instead of $C$
is simply that the syntaxes of values $v$ and evaluation contexts $E$
are not closed under substitution of arbitrary terms for
variables---they assume that variables are values---and we want
proposition well-formedness to be preserved under arbitrary term
substitutions.  All this means, practically speaking, is that
something like $x\,\hole : \tau \eto \tau'$ cannot hold categorically,
but only in a context where $x\in\OVal$ is also provable.

As explained in Section~\ref{sec:core-rules}, along with these new
atomic propositions, we will also make use of various first-order
theorems about them, which are provable straightforwardly in the
meta-logic without requiring any stepwise reasoning.  For example,
\[
\infer{
\con \ts e_1 \stepzero e_2 \ \lor\  e_2 \stepzero e_1
}{
\con \ts e\stepone e_1 \quad
\con \ts e\stepone e_2
}
\]
and
\[
\infer{
\con \ts E[e_1] \ciuleq E[e_2]
}{
\con \ts E : \tau \eto \tau' \quad
\con \ts e_1 : \tau \quad
\con \ts e_2 : \tau \quad
\con \ts e_1 \ciuleq e_2
}
\]
See the proofs in subsequent sections for more examples.

Finally, we will make use of some additional notation, which is
definable in terms of the atomic propositions we have introduced:
\[
\begin{array}{r@{\quad}c@{\quad}l}
e \valof \tau &\defeq& e : \tau ~\land~ e \in \OVal
\\[2pt]
e_1 \termto e_2 &\defeq& e_1 \evalsto e_2 ~\land~ e_2 \in \OVal
\\[2pt]
e_1 \termzero e_2 &\defeq& e_1 \stepzero e_2 ~\land~ e_2 \in \OVal
\\[2pt]
R : \TRel{\tau_1}{\tau_2} &\defeq& \forall x_1,x_2.\  (x_1,x_2) \in
R ~\limp~ x_1 : \tau_1 ~\land~ x_2 : \tau_2 
\\[2pt]
R : \VRel{\tau_1}{\tau_2} &\defeq& \forall x_1,x_2.\ (x_1,x_2) \in
R ~\limp~ x_1\valof \tau_1 ~\land~ x_2 \valof \tau_2 
\\[2pt]
(x_1 : \tau_1, x_2 : \tau_2).\ P &\defeq& (x_1,x_2).\ x_1 : \tau_1 ~\land~ x_2 : \tau_2 ~\land~ P
\\[2pt]
(x_1 \valof \tau_1, x_2 \valof \tau_2).\ P &\defeq& (x_1,x_2).\ x_1 \valof \tau_1 ~\land~ x_2 \valof \tau_2 ~\land~ P
\end{array}
\]

\begin{figure}
\begin{myfig}
\small
\[
\begin{array}{@{}r@{~~\defeq~~}l}
\Vrel{\alpha}\rho & R,\ \ \mbox{where}\ \rho(\alpha) = (\tau_1,\tau_2,R)\\
\Vrel{\tau_b}\rho & (x_1\valof\tau_b,x_2 \valof\tau_b).\ x_1 = x_2,\ \ \mbox{where}\ \tau_b\in\{\tunit,\tint,\tbool\}\\
\Vrel{\tpair{\tau'}{\tau''}}\rho &
  \begin{stackTL}
    (x_1\valof\rho_1(\tpair{\tau'}{\tau''}),x_2\valof\rho_2(\tpair{\tau'}{\tau''})).\\
    \quad\exists x_1',x_1'',x_2',x_2''.\ x_1 = \epair{x_1'}{x_1''} \ \land\ 
          x_2 = \epair{x_2'}{x_2''} \ \land\\ \quad \qquad \qquad \qquad \ \ 
(x_1',x_2') \in \Vrel{\tau'}\rho \ \land\  (x_1'',x_2'') \in \Vrel{\tau''}\rho
  \end{stackTL}\\
\Vrel{\tsum{\tau'}{\tau''}}\rho &
  \begin{stackTL}
    (x_1\valof\rho_1(\tsum{\tau'}{\tau''}),x_2\valof\rho_2(\tsum{\tau'}{\tau''})).\\
    \quad (\exists x_1',x_2'.\ x_1 = \einl{x_1'} \ \land\  x_2 = \einl{x_2'} \ \land\  (x_1',x_2') \in \Vrel{\tau'}\rho) \ \ \lor \\
    \quad (\exists x_1'',x_2''.\ x_1 = \einr{x_1''} \ \land\  x_2 = \einr{x_2''} \ \land\  (x_1'',x_2'') \in \Vrel{\tau''}\rho)) \\
  \end{stackTL}\\
\Vrel{\tfun{\tau'}{\tau''}}\rho &
  \begin{stackTL}
    (x_1\valof\rho_1(\tfun{\tau'}{\tau''}),x_2\valof\rho_2(\tfun{\tau'}{\tau''})). \\
    \quad \forall y_1,y_2.\ (y_1,y_2) \in \Vrel{\tau'}\rho \limp
              (x_1 y_1, x_2 y_2) \in \Crel{\tau''}\rho
  \end{stackTL}\\
\Vrel{\tall{\alpha}{\tau}}\rho &
  \begin{stackTL}
    (x_1\valof\rho_1(\tall{\alpha}{\tau}),x_2\valof\rho_2(\tall{\alpha}{\tau})). \\
    \quad \forall\alpha_1,\alpha_2.\ \forall r.\ r:\vrel{\alpha_1}{\alpha_2} \limp (\etapp{x_1}{\alpha_1},\etapp{x_2}{\alpha_2}) \in \Crel{\tau}\rho,\alpha\mapsto(\alpha_1,\alpha_2,r)
  \end{stackTL}\\
\Vrel{\texist{\alpha}{\tau}}\rho &
  \begin{stackTL}
    (x_1\valof\rho_1(\texist{\alpha}{\tau}),x_2\valof\rho_2(\texist{\alpha}{\tau})). \\
    \quad \exists\alpha_1,\alpha_2,y_1,y_2.\ \exists r.\ r:\vrel{\alpha_1}{\alpha_2} ~\land~\\ 
    \quad\quad  x_1 = \epack{\alpha_1}{y_1}{\texist{\alpha}{\rho_1\tau}} \ \land\ 
            x_2 = \epack{\alpha_2}{y_2}{\texist{\alpha}{\rho_2\tau}} \ \land\ \\
  \quad\quad(y_1,y_2) \in \Vrel{\tau}\rho,\alpha\mapsto(\alpha_1,\alpha_2,r)
  \end{stackTL}\\
\Vrel{\trec{\alpha}{\tau}}\rho &
  \begin{stackTL}
    \mu r.(x_1\valof\rho_1(\trec{\alpha}{\tau}),x_2\valof\rho_2(\trec{\alpha}{\tau})).\\
    \quad \exists y_1,y_2.\ x_1 = \efold{y_1} \ \land\  x_2 = \efold{y_2} \ \land \ \\ \quad \quad
     \lift(y_1,y_2) \in \Vrel{\tau}\rho,\alpha\mapsto(\rho_1(\trec{\alpha}{\tau}),\rho_2(\trec{\alpha}{\tau}),r)
  \end{stackTL}\\
\Crel{\tau}\rho & 
  \begin{stackTL}
    \mu r.(t_1:\rho_1\tau,t_2:\rho_2\tau). \\ \quad (\forall x_1.\ t_1 \termzero x_1 \limp \exists x_2.\ x_2 \ciuleq t_2 \land (x_1,x_2) \in \Vrel{\tau}{\rho})\ \land\ \\ \quad (\forall t_1'.\ t_1 \stepone t_1' \limp \lift(t_1',t_2) \in r)
  \end{stackTL}\\
\end{array}
\]
\mycaption{Syntactic Logical Relation for $\fmu$}
\label{fig:logrel}
\end{myfig}
\end{figure}

\subsection{Logical Relation}

Figure~\ref{fig:logrel} defines two logical relations for $\fmu$, one
for values ($\Vrel{\tau}{\rho}$) and one for terms
($\Crel{\tau}{\rho}$).  These are syntactic \thelogic{} relations,
defined by induction on~$\tau$.  Here, $\rho$ is assumed to be a
syntactic relational interpretation of the free type variables of
$\tau$, \ie a mapping from each $\alpha\in\mathrm{FV}(\tau)$ to a
triple $(\tau_1,\tau_2,R)$ such that $R:\VRel{\tau_1}{\tau_2}$.  We
write $\rho_i$ to mean the type substitution mapping each $\alpha$ to
the corresponding $\tau_i$.  Thus, it is trivial to prove that
$\Vrel{\tau}{\rho}:\vrel{\rho_1\tau}{\rho_2\tau}$ and
$\Crel{\tau}{\rho}:\trel{\rho_1\tau}{\rho_2\tau}$.  Except for the
last two cases ($\Vrel{\mu\alpha.\tau}{\rho}$ and
$\Crel{\tau}{\rho}$), the definition of the logical relation is
entirely straightforward, following Plotkin and
Abadi~\cite{plotkin-abadi-93}, with each type constructor being
modeled by its corresponding logical connective via the Curry-Howard
isomorphism.

First, let us consider $\Vrel{\mu\alpha.\tau}{\rho}$.  The basic idea
here is to give the relational interpretation of a recursive type
using a recursive relation $\mu r.R$.  Recall, though, that
references to $r$ in $R$ must only appear under ``later''
propositions.  Thus, we have that $\efold{v_1}$ and $\efold{v_2}$
are related by $\Vrel{\mu\alpha.\tau}{\rho}$ ``now'' iff $v_1$ and
$v_2$ are related by $\Vrel{\tau}\rho,\alpha\,{\mapsto}\,(\ldots,\Vrel{\mu\alpha.\tau}\rho) = \Vrel{\tau[\mu\alpha.\tau/\alpha]}{\rho}$ ``later''.

Next, consider $\Crel{\tau}{\rho}$.  Intuitively, we would like to say
that two terms $e_1$ and $e_2$ are related if, whenever $e_1$
evaluates to some value $v_1$, we have that $e_2$ also evaluates to
some value $v_2$ such that $(v_1,v_2) \in \Vrel{\tau}{\rho}$.  In
fact, in the case that $e_1$ evaluates to $v_1$ without incurring any
\cd{unroll}-\cd{roll} reductions (\ie when $e_1 \termzero v_1$), the
definition of $\Crel{\tau}{\rho}$ \emph{almost} says this---the only
difference is that instead of saying ``$e_2$ evaluates to some value
$v_2$ such that\ldots'', it says that ``$e_2$ is ciu-approximated by
some value $v_2$ such that\ldots'' Of course, by definition of ciu
approximation, this also implies that $e_2$ terminates, but it is
somewhat more liberal in that it does not require the value that $e_2$
produces to be directly related to $v_1$ by $\Vrel{\tau}\rho$.  This
extra freedom is not strictly necessary if we just want to define a
logical relation that is \emph{sound} w.r.t.\ contextual
approximation---as we did in the previous version of this
paper~\cite{dreyer+:lics09}---but it is key to ensuring
\emph{completeness} (see Theorems~\ref{thm:ciu-transitivity}
and~\ref{thm:ciu-log} in Section~\ref{sec:sound-complete}).  An
alternative approach to ensuring completeness would be to employ
$\top\top$-closure, as Pitts does~\cite{pitts:attapl}.  We discuss
this alternative in Section~\ref{sec:related}.

However, in the case that the evaluation of $e_1$ incurs an
\cd{unroll}-\cd{roll} reduction, the interpretation of recursive types
forces us to require something still weaker.  Specifically, in order
to prove that the logical relation is sound with respect to contextual
approximation, we must prove that it is \emph{compatible} in the sense
of Pitts~\cite{pitts:attapl}.  Compatibility for \cd{unroll} demands
that if $\eroll{v_1}$ and $\eroll{v_2}$ are logically related, then
$\eunroll{(\eroll{v_1})}$ and $\eunroll{(\eroll{v_2})}$ are related,
too.  By definition of $\Vrel{\mu\alpha.\tau}{\rho}$, knowing
$\eroll{v_1}$ and $\eroll{v_2}$ are related only tells us that $v_1$
and $v_2$ are related ``later''.  We need to be able to derive from
that that $\eunroll{(\eroll{v_1})}$ and $\eunroll{(\eroll{v_2})}$ are
related ``now''.  Thus, in defining whether
$(e_1,e_2)\in\Crel{\tau}{\rho}$, in the case that $e_1$ makes an
\cd{unroll}-\cd{roll} reduction (\ie $e_1 \stepone e_1'$), we only
require that $e_1'$ and $e_2$ be related \emph{later} (\ie
$\later(e_1',e_2)\in\Crel{\tau}\rho$).

For the reader who is familiar with prior work on step-indexed models
and logical relations, our formulation here may seem familiar and yet
somewhat unusual.  Our use of the later operator corresponds to where
one would ``go down a step'' in the construction of a step-indexed
model.  However, in prior work, step-indexed models typically go down
a step \emph{everywhere} (\ie in every case of the logical relation),
not just in one or two places, and ``count'' every step, not just
\cd{unroll}-\cd{roll} reductions.  If one is working with
\emph{equi-recursive} types, this may be the only option, but here we
are working with \emph{iso-recursive} types, and our present
formulation serves to isolate the use of the later operator to the few
places where it is absolutely needed.  While we do not believe there
is a fundamental difference between what one can prove using this
logical relation vs.\ previous accounts, our formulation enables more
felicitous statements of certain properties, such as the
extensionality principle for functions (see discussion of
Rule~\rul{funext} below).

Finally, it is worth noting that, like step-indexed models,
\thelogic{} imposes no ``admissibility'' requirement on candidate
relations.  Intuitively, the reason admissibility is unnecessary is
that it is an infinitary property.  In \thelogic{}, we only ever
reason about finitary properties, \ie propositions that hold true in
the ``current'' world; we do not even have the ability (within the
logic) to talk about truth in all worlds.

\begin{figure}[!t]
\begin{myfig}
\small
\[
\inferno{val}{
\con \ts \crelated{e_1}{e_2}{\tau}{\rho}
}{
\con \ts \vrelated{e_1}{e_2}{\tau}{\rho}
}
\qquad
\inferno{weak-$\later$}{
\con \ts \later P
}{
\earlier\con \ts P
}
\]
\betweenrules
\[
\inferno{exp}{
\ttwf{\con}{\erelated{e_1}{e_2}{\tau}{\rho}}
}{
\begin{array}{c}
\con \ts e_1 : \rho_1\tau \\
\ttwf{\con}{e_1 \evalsto e_1'} \quad
\ttwf{\con}{\erelated{e_1'}{e_2}{\tau}{\rho}}
\end{array}
}
\quad
\inferno{exp-$\later$}{
\ttwf{\con}{\erelated{e_1}{e_2}{\tau}{\rho}}
}{
\begin{array}{c}
\con \ts e_1 : \rho_1 \tau \quad
\con \ts e_2 : \rho_2 \tau \\
\ttwf{\con}{e_1 \stepone e_1'} \quad
\ttwf{\con}{\later\erelated{e_1'}{e_2}{\tau}{\rho}}
\end{array}
}
\]
\betweenrules
\[
\inferno{red}{
\ttwf{\con}{\erelated{e_1}{e_2}{\tau}{\rho}}
}{
\ttwf{\con}{e_1' \stepzero e_1} \quad
\ttwf{\con}{\erelated{e_1'}{e_2}{\tau}{\rho}}
}
\quad
\inferno{ciu}{
\ttwf{\con}{\erelated{e_1}{e_2}{\tau}{\rho}}
}{
\ttwf{\con}{\erelated{e_1}{e_2'}{\tau}{\rho}} \quad
\con \ts e_2' \ciuleq e_2
}
\]
\betweenrules
\[
\inferno{bind}{
\ttwf{\con}{\erelated{E[e_1]}{f}{\tau'}{\rho'}}
}{
\begin{array}{c}
\con \ts E : \rho_1\tau \eto \rho'_1\tau' \quad
\con \ts f : \rho_2'\tau' \quad
\ttwf{\con}{\erelated{e_1}{e_2}{\tau}{\rho}} \\[\bp]
\ttwf{\con,x_1,x_2,\vrelated{x_1}{x_2}{\tau}{\rho},e_1 \evalsto x_1, x_2 \ciuleq e_2}{\erelated{E[x_1]}{f}{\tau'}{\rho'}}
\end{array}
}
\]
\betweenrules
\[
\inferno{bind2}{
\ttwf{\con}{\erelated{E_1[e_1]}{E_2[e_2]}{\tau'}{\rho'}}
}{
\begin{array}{c}
\con \ts E_1 : \rho_1\tau \eto \rho'_1\tau' \quad
\con \ts E_2 : \rho_2\tau \eto \rho_2'\tau' \quad
\ttwf{\con}{\erelated{e_1}{e_2}{\tau}{\rho}} \\[\bp]
\ttwf{\con,x_1,x_2,\vrelated{x_1}{x_2}{\tau}{\rho},e_1 \evalsto x_1, x_2 \ciuleq e_2}{\erelated{E_1[x_1]}{E_2[x_2]}{\tau'}{\rho'}}
\end{array}
}
\]
\betweenrules
\[
\inferno{app}{
\ttwf{\con}{\erelated{\eapp{f_1}{e_1}}{\eapp{f_2}{e_2}}{\tau''}{\rho}}
}{
\ttwf{\con}{\erelated{f_1}{f_2}{\tfun{\tau'}{\tau''}}{\rho}} \quad
\ttwf{\con}{\erelated{e_1}{e_2}{\tau'}{\rho}}
}
\]
\betweenrules
\[
\inferno{unroll}{
\ttwf{\con}{\erelated{\eunroll{e_1}}{\eunroll{e_2}}{\tau[\trec{\al}{\tau}/\al]}{\rho}}
}{
\ttwf{\con}{\erelated{e_1}{e_2}{\trec{\al}{\tau}}{\rho}}
}
\]
\betweenrules
\[
\inferno{funext}{
\ttwf{\con}{\vrelated{e_1}{e_2}{\tfun{\tau'}{\tau''}}{\rho}}
}{
\begin{array}{c}
\con \ts e_1 \valof \rho_1(\tfun{\tau'}{\tau''}) \quad
\con \ts e_2 \valof \rho_2(\tfun{\tau'}{\tau''}) \\[\bp]
\con, x_1, x_2, \vrelated{x_1}{x_2}{\tau'}{\rho} \ts \crelated{e_1x_1}{e_2x_2}{\tau''}{\rho}
\end{array}
}
\]
\betweenrules
\[
\inferno{fix}{
\con \ts
(F_1,F_2) \in \Vrel{\tfun{\tau'}{\tau''}}{\rho}
}{
\begin{array}{c}
F_i = \efix{f}{x_i}{e_i} \quad% F_2 = \efix{f}{x_2}{e_2} \\
\con \ts F_1 : \rho_1(\tfun{\tau'}{\tau''}) \quad
\con \ts F_2 : \rho_2(\tfun{\tau'}{\tau''}) \\[\bp]
\earlier\con,x_1,x_2,(x_1,x_2)\in\Vrel{\tau'}{\rho},(F_1,F_2) \in \Vrel{\tfun{\tau'}{\tau''}}{\rho}\ts (e_1[F_1/f],e_2[F_2/f]) \in \Crel{\tau''}{\rho}
\end{array}
}
\]
\mycaption{Some Useful Derivable Rules}
\label{fig:derivable}
\end{myfig}
\end{figure}

\subsection{Derivable Rules}
\label{sec:derivable}

Figure~\ref{fig:derivable} shows a number of useful inference rules
that are derivable in the logic.  To be clear, by ``derivable'' we
mean that the proofs of these rules' soundness (given below in
Section~\ref{sec:proofs-derivable}) is done just using the inference
rules we have established so far, without needing to appeal directly
to the model and perform stepwise reasoning.  In all these rules, we
assume implicitly that all propositions are well-formed.  For the
rules concerning $\Vrel{\tau}{\rho}$ and $\Crel{\tau}{\rho}$, we
assume that $\rho$ binds the free variables of $\tau$ and maps them to
triples $(\tau_1,\tau_2,R)$, where $R:\vrel{\tau_1}{\tau_2}$ is
provable in the ambient context.

Rule~\rul{val} says that $\Crel{\tau}{\rho}$ contains
$\Vrel{\tau}{\rho}$.  This rule is so fundamental and ubiquitously
useful that we will often elide mention of it in our proofs.

Rule~\rul{weak-$\later$} is a weakening property that is easy to derive
from the distributivity laws for the $\later$ operator.  
The rule employs an $\earlier$ operator (pronounced ``earlier'') on
propositions/contexts, defined as follows:
\[
\begin{array}{r@{~~}c@{~~}l}
\earlier (\vcon;\rcon;\seq{P}) &\defeq&
\vcon;\rcon;\seq{\earlier P}
\\[2pt]
\earlier (\later P) &\defeq& P 
\\[2pt]
\earlier P &\defeq& P  \quad
\mbox{(if $P \neq \later P'$)}
\end{array}
\]
This $\earlier$ operator has the effect of ``un-$\later$-ing'' (\ie
stripping the $\later$ off of) any $\later P$ hypotheses in the
context.  Note that this is purely a shallow syntactic operation; it
does not un-$\later$ any hypotheses that are propositionally
equivalent to some $\later P$ but not syntactically of that form.
(The reader may wonder why we define $\earlier$ in this syntactic way
instead of building it in as a primitive modality with the seemingly
natural interpretation $\interp{\earlier P}{n}{\delta} =
\interp{P}{(n+1)}{\delta}$.  The trouble is that this interpretation
is not well-founded, since it defines the meaning of $\earlier P$ in
terms of the meaning of $P$ at a \emph{higher} step level.  And
indeed, our syntactic $\earlier$ does not satisfy this
interpretation.)

Consequently, Rule~\rul{weak-$\later$} says that if we want to show
$P$ is true later, given some assumptions that are true now, and
others that are true later, then we can just prove that $P$ is true
now given that all the assumptions are true now.  This is a weakening
property because, applying the rule backwards, we forget the fact that
some of the hypotheses in $\con$ (namely, those that are \emph{not} of
the form $\later P$) are true at an earlier world than the others.

The \rul{weak-$\later$} rule is particularly useful in conjunction
with the \rul{l\"ob} rule.  Specifically, thanks to the \rul{l\"ob}
rule, a frequently effective approach to proving two terms $e_1$ and
$e_2$ related is to assume inductively that they are related
\emph{later} and then prove that they are related now.  Eventually, we
may reduce our proof goal (via, \eg Rule~\rul{exp-$\later$}, explained
below) to showing that two other terms $e_1'$ and $e_2'$ are related
\emph{later}.  At that point, Rule~\rul{weak-$\later$} allows us to
un-$\later$ both our new proof goal (relatedness of $e_1'$ and $e_2'$)
and our original \rul{l\"ob}-inductive hypothesis (relatedness of
$e_1$ and $e_2$) simultaneously.  We will see an instance of this
proof pattern in the example in Section~\ref{sec:minimal-invariance}.

The next four rules in Figure~\ref{fig:derivable} allow one to prove
that two terms $e_1$ and $e_2$ are related by converting one of the
terms to something else.  Rule~\rul{exp} (closure of the logical
relation under expansion) allows one to reduce $e_1$ to some $e_1'$
according to the $\stepmany$ relation and then show that $e_1'$ is
related to $e_2$.  Rule~\rul{red} (closure of the logical relation
under $\stepzero$ reduction) allows one to expand $e_1$ to some $e_1'$
according to the $\stepzero$ relation and then show that $e_1'$ is
related to $e_2$.  Rule~\rul{ciu} allows one to replace $e_2$ with
some $e_2'$ that ciu-approximates it, and then show that $e_1$ is
related to $e_2'$.  Rule~\rul{exp-$\later$} is similar to
Rule~\rul{exp}, but addresses the case when $e_1$ incurs an
\cd{unroll}-\cd{roll} reduction on the way to $e_1'$.  In this case,
unfolding the definition of $\Crel{\tau}{\rho}$, all we have to show
is that $e_1'$ and $e_2$ are related \emph{later}.

The aforementioned rules are all useful when we know what the terms in
question reduce/expand to.  Rule~\rul{bind} is important because it
handles the case when a term is ``stuck''.  For instance, suppose we
want to show that $e$ and $f$ are related, where $e$ is of the form
$E[e_1]$ (\ie $e_1$ is in evaluation position in $e$, and $E$ is the
evaluation context surrounding it).  Perhaps $e_1$ is something like
$y_1(v_1)$, in which case there is no way to reduce it.  However, if
we can prove that $y_1(v_1)$ is logically related to some other
expression $e_2$, then there are two cases to consider.  In the case
that they both terminate, we can assume that there are some values
$x_1$ and $x_2$ such that $e_1$ evaluates to $x_1$, $e_2$ is
ciu-approximated by $x_2$, and $x_1$ and $x_2$ are related by
$\Vrel{\tau}{\rho}$, and the goal is reduced to showing that $E[x_1]$
is related to $f$.  In the case that $e_1$ diverges, there is nothing
to show, since $E[e_1]$ will diverge, too.

The \rul{bind} rule may seem at first glance a bit peculiar in that
the term $e_2$ does not necessarily have any relationship to $f$, and
the variable $x_2$ does not appear anywhere on the r.h.s.\ of the last
premise.  This peculiarity is a consequence of the rule being as
general as possible.  In the specific (if common) case that $f$ is in
fact of the form $E_2[e_2]$ (\ie that $e_2$ is in evaluation position
in $f$), an easy corollary of Rules~\rul{bind} and~\rul{ciu} is
Rule~\rul{bind2}.  In addition to being more intuitive, this more
symmetric-looking variant of the \rul{bind} rule is very useful in
deriving \emph{compatibility} properties~\cite{pitts:attapl}, such as
Rules~\rul{app} and \rul{unroll}; these compatibility properties are
necessary in order to establish that the logical relation is a
precongruence (and hence contained in contextual approximation), and
Rule~\rul{bind2} helps to reduce the derivations of these properties
to the case where the $e$'s and $f$'s are values.  Rule~\rul{bind2}
does not subsume Rule~\rul{bind}, however: the general and distinctly
asymmetric nature of the original Rule~\rul{bind} renders it suitable
for reasoning about logical approximation in cases where the more
symmetric Rule~\rul{bind2} does not apply---for instance, see the
proof of the ``free theorem'' example in
Section~\ref{sec:free-theorem}.

Rule~\rul{funext} demonstrates a clean extensionality property for
function values, which was one of our key motivations for LSLR in the
first place.  (The property does not hold for arbitrary terms in our
call-by-value semantics.)  It is worth noting that, in prior
step-indexed models, this extensionality property is not quite so
clean to state.  For example, if one were to encode Ahmed's
relation~\cite{ahmed-2006} in our logic directly, the assumption
$(x_1,x_2)\in\Vrel{\tau'}{\rho}$ would have to be $\later$'d.  The key
to our cleaner formulation is simply that we confine the use of
$\later$ in $\Vrel{\tau}\rho$ to the case when $\tau$ is a recursive
type.  Thus, in particular, one need not mention $\later$ when
reasoning purely about functions and $\beta$-reduction.

Finally, Rule~\rul{fix} gives the rule for recursive functions,
which are encodable in a well-known way in terms of recursive types.
We formalize the encoding as follows:
\[
\begin{array}{r@{~}c@{~}l}
\efix{f}{x}{e} &\defeq& \eunf{y}(\eunfold{v})\,v\,y \\[2pt]
\quad\mbox{where}\ v&=&\begin{stackTL}
\efold{(\eunf{z}(\eunf{f}\eunf{x}e)(\eunf{y}(\eunfold{z})\,z\,y))} \\
\mbox{for $y,z\not\in\FV(e)$}
\end{stackTL}
\end{array}
\]
This encoding has the property that if $F = \efix{f}{x}{e}$, then
$F(v) \stepone e[F/f,v/x]$.  Consequently, to show two recursive
functions related, we may \rul{l\"ob}-inductively assume they are
related while proving that their bodies are related.  (For the proof
that the bodies are related, we may also un-$\later$ any other
$\later$ hypotheses in the ambient context $\con$.)  The implicit
use of \rul{l\"ob} induction in this rule gives it a distinctively
coinductive flavor.

\subsection{Proofs of Derivability}
\label{sec:proofs-derivable}

In this section, we show how to derive the rules in
Figure~\ref{fig:derivable}.

\begin{prop}[Type Substitution]
\label{prop:typesubst}
\label{prop:logrel:typesubst}\hfill
\begin{enumerate}[\em(1)]
\item
$\Vrel{\tau[\sigma/\alpha]}\!\rho =
\Vrel{\tau}\!\rho,\alpha{\,\mapsto\,}(\rho_1\sigma,\rho_2\sigma,\Vrel{\sigma}\!\rho)$.
\item
$\Crel{\tau[\sigma/\alpha]}\!\rho =
\Crel{\tau}\!\rho,\alpha{\,\mapsto\,}(\rho_1\sigma,\rho_2\sigma,\Vrel{\sigma}\!\rho)$.
\end{enumerate}
\end{prop}

\proof
By straightforward induction on the structure of $\tau$.
\qed

\begin{prop}
Rule~\rul{val} is derivable.
\end{prop}

\proof
Immediate, since $\ciuleq$ is reflexive.
\qed

\begin{prop}
Rule~\rul{weak-$\later$} is derivable.
\end{prop}

\proof Suppose $\con = \vcon;\rcon;\pcon$.  Then, $\earlier\con \ts P$
implies $\vcon;\rcon;\empctx\ts(\bigwedge_{Q\in\pcon}\,\earlier Q) \limp P$.  By
Rule~\rul{mono} and the distributivity axioms,
$\vcon;\rcon;\empctx\ts(\bigwedge_{Q\in\pcon}\, \later\earlier Q) \limp \later P$.
Since $Q \limp \later\earlier Q$, we have
$\vcon;\rcon;\empctx\ts(\bigwedge_{Q\in\pcon} Q) \limp \later P$, and
thus $\con \ts \later P$.
\qed

\begin{prop}
\label{prop:red}
Rule~\rul{red} is derivable.
\end{prop}

\proof First, suppose that $e_1 \termzero x_1$ for some value $x_1$.
Then, $e_1' \stepzero e_1$ implies that $e_1' \termzero x_1$ as well,
and the rest follows immediately from $(e_1',e_2) \in
\Crel{\tau}\rho$.

Second, suppose that $e_1 \stepone t_1$ for some term $t_1$.  Then,
$e_1' \stepzero e_1$ implies that $e_1' \stepone t_1$ as well, so
again the rest follows immediately from $(e_1',e_2) \in
\Crel{\tau}\rho$.
\qed

\begin{prop}
\label{prop:exp-zero}
Rule~\rul{exp} is derivable given the additional premise that $\con \ts e_1 \stepzero e_1'$.
\end{prop}

\proof
The proof is very similar to the proof of Rule~\rul{red}.  The key
bits are: (1) if $e_1 \termzero x_1$ and $e_1 \stepzero e_1'$, then
$e_1' \termzero x_1$ by determinacy of reduction, and (2)
if $e_1 \stepone t_1$ and $e_1 \stepzero e_1'$, then $e_1' \stepone t_1$,
again by determinacy of reduction.
\qed

\begin{prop}
Rule~\rul{exp-$\later$} is derivable.
\end{prop}

\proof
First, suppose that $e_1 \termzero x_1$ for some value $x_1$.
Then, $e_1 \stepone e_1'$ yields a contradiction.

Second, suppose that $e_1 \stepone t_1$ for some term $t_1$.  Then,
since $e_1 \stepone e_1'$, we have by determinacy of reduction that
either $e_1' \stepzero t_1$ or $t_1 \stepzero e_1'$.  Thus, by either
Proposition~\ref{prop:red} or~\ref{prop:exp-zero}, $\later(e_1',e_2)
\in \Crel{\tau}\rho$ implies $\later(t_1,e_2) \in \Crel{\tau}\rho$,
which is what we needed to show.  \qed

\begin{prop}
Rule~\rul{exp} is derivable.
\end{prop}

\proof Assume the premises of Rule~\rul{exp}.  We will prove the
following proposition and then instantiate $t_1$ with $e_1$ to obtain
the desired result.
\[
\forall t_1.\ (t_1 : \rho_1\tau \land t_1 \stepmany e_1') \limp
(t_1,e_2) \in \Crel{\tau}\rho
\]
The proof is by \emph{\rul{l\"ob} induction}, \ie we use the
\rul{l\"ob} rule to assume the above proposition is true ``later''
(under a $\later$ modality) and then prove it true ``now''.  So assume
$t_1 : \rho_1\tau$ and $t_1 \stepmany e_1'$, and we want to prove
$(t_1,e_2) \in \Crel{\tau}\rho$.  It is thus either the case that $t_1
\stepzero e_1'$ or that there exists $t_1'$ such that $t_1 \stepone
t_1' \stepmany e_1'$.  In the former case, the result follows by
Proposition~\ref{prop:exp-zero} and the assumption $(e_1',e_2) \in
\Crel{\tau}\rho$.  In the latter case we have, by the
\rul{l\"ob}-inductive hypothesis (\ie the $\later$-ed version of our
original goal) together with the distributivity of $\later$ over
$\forall$ and $\limp$, that $\later (t_1' : \rho_1\tau \land t_1'
\stepmany e_1') \limp \later (t_1',e_2) \in \Crel{\tau}\rho$.  We
already know that $t_1' \stepmany e_1'$, and $t_1' :\rho_1\tau$
follows by type preservation, so by Rule~\rul{mono}, we have that
$\later(t_1',e_2) \in \Crel{\tau}\rho$.  The result then follows from
$t_1 \stepone t_1'$ and Rule~\rul{exp-$\later$}.  \qed

\begin{prop}
Rule~\rul{ciu} is derivable.
\end{prop}

\proof
As for Rule~\rul{exp}, the proof here is by \rul{l\"ob} induction.
Given the premises of Rule~\rul{ciu}, we prove the following
and then instantiate $t_1$ to $e_1$:
\[
\forall t_1.\ (t_1,e_2') \in \Crel{\tau}\rho \limp (t_1,e_2) \in \Crel{\tau}\rho
\]
Assume this is true later, and we proceed to prove it now.  So assume
$(t_1,e_2') \in \Crel{\tau}\rho$, and we want to prove $(t_1,e_2) \in
\Crel{\tau}\rho$.

First, suppose $t_1 \termzero x_1$.  Then, there exists $x_2$ such
that $(x_1,x_2) \in \Vrel{\tau}\rho$ and $x_2 \ciuleq e_2'$.
Since by assumption $e_2' \ciuleq e_2$ and $\ciuleq$ is transitive,
we have that $x_2 \ciuleq e_2$, so we are done.

Second, suppose $t_1 \stepone t_1'$.  Then,
$\later(t_1',e_2')\in\Crel{\tau}\rho$, so by the \rul{l\"ob}-inductive
hypothesis, $\later(t_1',e_2)\in\Crel{\tau}\rho$.
\qed

\begin{prop}
Rule~\rul{bind} is derivable.
\end{prop}

\proof
Define $P(t_1)$ to be the proposition:
\[
\begin{stackTL}
\forall x_1,x_2.\ ((x_1,x_2)\in\Vrel{\tau}{\rho} \land
                   t_1 \evalsto x_1 \land x_2 \ciuleq e_2)
\limp (E[x_1],f) \in \Crel{\tau'}{\rho'}
\end{stackTL}
\]
We want to prove that
\[
\begin{stackTL}
\forall t_1.\ ((t_1,e_2)\in\Crel{\tau}\rho \land P(t_1)) \limp (E[t_1],f) \in \Crel{\tau'}{\rho'}
\end{stackTL}
\]
By the \rul{l\"ob} rule, we assume this proposition is true later and
proceed to prove it now.  So assume that
$(t_1,e_2)\in\Crel{\tau}{\rho}$ and $P(t_1)$, and we want to prove
$(E[t_1],f)\in \Crel{\tau'}{\rho'}$.

First, suppose that $E[t_1] \termzero x_1$ for some $x_1$.
Then, it must be the case that $t_1 \termzero y_1$ for some $y_1$,
and also that $E[t_1] \stepzero E[y_1] \termzero x_1$.
Since $(t_1,e_2)\in\Crel{\tau}{\rho}$, we know there exists some
$y_2$ such that $y_2 \ciuleq e_2$ and $(y_1,y_2)\in\Vrel{\tau}{\rho}$.
Thus, by $P(t_1)$, we know that $(E[y_1],f)\in\Crel{\tau'}{\rho'}$.
Then, by Rule~\rul{exp}, $(E[t_1],f)\in\Crel{\tau'}{\rho'}$.

Second, suppose that $E[t_1] \stepone t_1'$.  There are two cases:

\begin{cases}
\casen{1} \zilch\\
There exists $y_1$ such that $t_1 \termzero y_1$, and
also that $E[t_1] \stepzero E[y_1] \stepone t_1'$.  The proof is identical
to the previous case shown above.

\casen{2} \zilch\\
There exists $u_1$ such that $t_1 \stepone u_1$, and
also that $E[t_1] \stepone E[u_1] \stepzero t_1'$.  Since
$(t_1,e_2)\in\Crel{\tau}{\rho}$, we know that
$\later(u_1,e_2)\in\Crel{\tau}{\rho}$.  Also, it is easy to show that
$P(t_1)$ implies $P(u_1)$.  Thus, by appealing to our \rul{l\"ob}-inductive
hypothesis, we have that $\later(E[u_1],f)\in\Crel{\tau'}{\rho'}$.
Finally, by Rule~\rul{red},
$\later(t_1',f)\in\Crel{\tau'}{\rho'}$.
\qed
\end{cases}

\begin{prop}
Rule~\rul{bind2} is derivable.
\end{prop}

\proof
By Rules~\rul{bind} and~\rul{ciu}, together with the fact that
$x_2 \ciuleq e_2$ implies $E_2[x_2] \ciuleq E_2[e_2]$, by part (4)
of Proposition~\ref{prop:ciu-props}.
\qed

\begin{prop}
Rule~\rul{app} is derivable.
\end{prop}

\proof By Rule~\rul{bind2}, using evaluation contexts $\hole\,e_1$ and
$\hole\,e_2$, the goal reduces to showing that
$(x_1\,e_1,x_2\,e_2)\in\Crel{\tau''}\rho$ under the assumption that
$(x_1,x_2)\in\Vrel{\tfun{\tau'}{\tau''}}\rho$.  By Rule~\rul{bind2}
again, this time using evaluation contexts $x_1\,\hole$ and
$x_2\,\hole$, the goal reduces to showing that
$(x_1\,y_1,x_2\,y_2)\in\Crel{\tau''}\rho$ under the assumption that
$(y_1,y_2)\in\Vrel{\tau'}\rho$.  The result then follows by unrolling
the definition of $\Vrel{\tfun{\tau'}{\tau''}}\rho$.  \qed

\begin{prop}
Rule~\rul{unroll} is derivable.
\end{prop}

\proof By Rule~\rul{bind2}, using the evaluation context
$\eunroll{\hole}$ on both sides, the goal reduces to showing that
$(\eunroll{x_1},\eunroll{x_2}) \in
\Crel{\tau[\mu\alpha.\tau/\alpha]}\rho$ under the assumption that
$(x_1,x_2) \in \Vrel{\mu\alpha.\tau}\rho$.  Unrolling the definition
of $\Vrel{\mu\alpha.\tau}\rho$, we have that $x_1 = \eroll{y_1}$, $x_2
= \eroll{y_2}$, and
$\later(y_1,y_2)\in\Vrel{\tau}\rho,\alpha\,{\mapsto}\,\Vrel{\mu\alpha.\tau}\rho$
for some $y_1$ and $y_2$.  By Proposition~\ref{prop:typesubst},
$\later(y_1,y_2)\in\Vrel{\tau[\mu\alpha.\tau/\alpha]}\rho$.  Also, we
have that $\eunroll{x_i} = \eunroll{(\eroll{y_i})} \stepone y_i$ (and
thus $y_i \ciuleq \eunroll{x_i}$ as well).  Thus, the desired result
follows directly by Rule~\rul{exp-$\later$} and Rule~\rul{ciu}.
\qed

\begin{prop}
Rule~\rul{funext} is derivable.
\end{prop}

\proof
Immediate, by unfolding the definition of $\Vrel{\tfun{\tau'}{\tau''}}\rho$.
\qed

\begin{prop}
Rule~\rul{fix} is derivable.
\end{prop}

\proof By straightforward combination of Rules~\rul{l\"ob},
\rul{funext}, \rul{weak-$\later$}, \rul{exp-$\later$}, and \rul{ciu},
given the fact that $F_i\,x_i \stepone e_i[F_i/f]$.  \qed

\subsection{Soundness and Completeness of the Logical Relation}
\label{sec:sound-complete}

We now state some key theorems concerning the logical relation, the
primary ones being that it is sound and complete w.r.t.\ contextual
approximation.

\begin{defi}[Logical Approximation] 
\label{def:logaprx} Let
  $\judg{\Gamma}{e_1}{\tau}$ and $\judg{\Gamma}{e_2}{\tau}$. \\
Suppose $\Gamma = {\alpha_1,\ldots,\alpha_n,x_1{\,:\,}\tau_1,\ldots,x_m{\,:\,}\tau_m}$.  Let 
\[
\begin{array}{@{}r@{~}c@{~}l}
\vcon&=& \alpha_1^1,\alpha_1^2,\ldots,\alpha_n^1,\alpha_n^2,
x_1^1,x_1^2,\ldots, x_m^1,x_m^2 \\[2pt]
\rcon&=& r_1,\ldots,r_n \\[2pt]
\rho &=& \alpha_1{\,\mapsto\,}(\alpha_1^1,\alpha_1^2,r_1),    
          \ldots, \alpha_n{\,\mapsto\,}(\alpha_n^1,\alpha_n^2,r_n) \\[2pt]
\pcon&=& r_1 : \VRel{\alpha_1^1}{\alpha_1^2},\ldots,r_n : \VRel{\alpha_n^1}{\alpha_n^2}, \\[2pt]
&& (x_1^1,x_1^2)\in\Vrel{\tau_1}\rho,
  \ldots, (x_m^1,x_m^2)\in\Vrel{\tau_m}\rho \\[2pt]
\gamma_j &=& x_1{\,\mapsto\,}x_1^j, 
   \ldots,x_m{\,\mapsto\,}x_m^j \ \mbox{(where $j \in \{1,2\}$)}\\
\end{array}
\]
Then
\[
\begin{array}{l}
\Gamma \ts e_1 \logaprx e_2 : \tau ~\defeq~
\vcon;\rcon;\pcon \ts
 (\rho_1\gamma_1e_1,\rho_2\gamma_2e_2) \in \Crel{\tau}\rho
\end{array}
\]
\end{defi}

\begin{thm}[Fundamental Theorem of Logical Relations]
\label{thm:parametricity}\hfill\\
If $\Gamma \ts e : \tau$ then $\Gamma \ts e \logaprx e : \tau$. 
\end{thm}

\proof By induction on typing derivations.  In the case when $e$ is a
variable, the goal follows directly from the hypotheses $\pcon$ in
Definition~\ref{def:logaprx}.  All of the other cases follow
immediately from the compatibility rules, which are all completely
straightforward to prove (in the style of Rule~\rul{app}).  The only
slightly interesting compatibility rule is Rule~\rul{unroll}, which we
proved in Section~\ref{sec:proofs-derivable}.  \qed

\begin{thm}[Adequacy]
\label{thm:adequacy}\hfill\\
If $\ \ts (e_1,e_2) \in \Crel{\tau}$ and $e_1 \termto$, then
$e_2 \termto$.
\end{thm}

\proof
Suppose $e_1 \termto v_1$.
Let $n$ be the number of \cd{unroll}-\cd{roll} reductions that occur
in the evaluation of $e_1$ to $v_1$.  It is easy to show by
induction on $n$, and by unfolding the definition of
$\Crel{\tau}$, that $\ \ts \later^n(v_1,e_2) \in
\Crel{\tau}$ (where $\later^n$ denotes $n$ applications of
the $\later$ modality).  Thus, $\ts \later^n(\exists x_2\valof\tau.\ x_2
\ciuleq e_2)$.

Appealing to the model, we have that $\forall k\geq 0.\ \interp{\later^n(\exists x_2\valof \tau.\ x_2\ciuleq e_2)}{k}{}$.
Choosing $k > n$, this means that there exists a value $v_2:\tau$ such
that $v_2 \ciuaprx e_2$.  Hence, $e_2 \termto$.  \qed

\begin{thm}[Logical Approximation $\limp$ Contextual Approximation]
\label{thm:log-ctx}
\label{thm:logrel:ctx-sound}\hfill\\
If
$\Gamma \ts e_1 \logaprx e_2 : \tau$, then $\Gamma \ts e_1 \ctxaprx e_2 : \tau$. 
\end{thm}

\proof Given a context $C : (\twf{\Gamma}{\tau}) \eto
(\twf{\Gamma'}{\tau'})$, we show that $\Gamma' \ts C[e_1] \logaprx
C[e_2] : \tau'$.  The proof of this part is by induction on the
context $C$, and as in the proof of the Fundamental Theorem, all of
the cases follow immediately from the compatibility rules.  Thus, if
$\Gamma'$ is empty, we know that $\ts (C[e_1],C[e_2]) \in
\Crel{\tau'}$.  Consequently, by Adequacy, we know that
$C[e_1]\termto$ implies $C[e_2]\termto$. \qed

\begin{thm}[Ciu-Transitivity of the Logical Relation]
\label{thm:ciu-transitivity} \hfill \\
If $\Gamma \ts e_1 \logaprx e_2' : \tau$ and $\Gamma \ts e_2' \ciuaprx e_2 : \tau$, then $\Gamma \ts e_1 \logaprx e_2 : \tau$.
\end{thm}

\proof Let $\vcon$, $\rcon$, $\pcon$, $\rho$, and $\gamma_j$ be as
defined in Definition~\ref{def:logaprx}.  From the second assumption,
it is easy to show by appeal to the model that $\vcon;\rcon;\pcon \ts
\rho_2\gamma_2e_2' \ciuleq \rho_2\gamma_2e_2$.  Thus, the result
follows immediately by Rule~\rul{ciu}.  \qed

\begin{thm}[Ciu Approximation $\limp$ Logical Approximation]
\label{thm:ciu-log} \hfill \\
If $\Gamma \ts e_1 \ciuaprx e_2 : \tau$, then $\Gamma \ts e_1 \logaprx e_2 : \tau$.
\end{thm}

\proof By the Fundamental Theorem of Logical Relations,
$\Gamma \ts e_1 \logaprx e_1 : \tau$.  The result then follows
directly by Theorem~\ref{thm:ciu-transitivity}.  \qed

\begin{cor}[$\ctxaprx ~\equiv~ \ciuaprx ~\equiv~ \logaprx$]
\label{thm:coincidence} \hfill \\
$\Gamma \vdash e_1 \ctxaprx e_2 : \tau$ iff $\Gamma \vdash e_1 \ciuaprx e_2 : \tau$ iff $\Gamma \vdash e_1 \logaprx e_2 : \tau$.
\end{cor}

\proof
By Theorems~\ref{thm:ctx-ciu}, \ref{thm:log-ctx} and \ref{thm:ciu-log}.
\qed

\subsection{Symmetric Version of the Logical Relation}

We have shown that our logical relation supports sound
\emph{inequational} reasoning about contextual approximation, but we
would like to support \emph{equational} reasoning as well.  Of course,
one can prove two terms equivalent by proving that each approximates
the other, but often this results in a tedious duplication of work.
Fortunately, we can define a symmetric version of our logical relation
directly in terms of the asymmetric one.

First, some notation: for a binary term relation $R$, let
$\reverse{R}$ denote $(t_2,t_1).(t_1,t_2)\in R$.  Also, let
$\reverse{\rho}$ denote the mapping with domain equal to that of
$\rho$ such that if $\rho(\alpha)=(\tau_1,\tau_2,R)$, then
$\reverse{\rho}(\alpha)=(\tau_2,\tau_1,\reverse{R})$.

Now, perhaps the most natural way of defining a symmetric version of
our logical relation would be to say that two terms/values are
symmetrically related if they are \emph{logically equivalent}, \ie
asymmetrically related (by $\Crel{\tau}$) in both directions.
Interestingly, this does not work.  In particular, there are a variety
of properties (described below) that we would like our symmetric
relation to enjoy, one of them being the property that
symmetrically-related function values $f_1$ and $f_2$ (of type
$\tau'\to\tau''$) are precisely those that map symmetrically-related
arguments (of type $\tau'$) to symmetrically-related results (of type
$\tau''$).  However, just knowing that $f_1$ and $f_2$ map
\emph{equivalent} arguments to \emph{equivalent} results does not
imply that they are equivalent themselves; to show equivalence, we
would need to establish relatedness of $f_1$ and $f_2$ in both
directions, which would at a minimum require that they map
$\Vrel{\tau'}$-related arguments (which are not necessarily
equivalent) to $\Crel{\tau''}$-related results.  Merely knowing how
$f_1$ and $f_2$ behave on \emph{equivalent} arguments is not enough
to establish that.

\begin{figure}[!t]
\[
\begin{array}{rcl}
\Vrelsym{\tau}{\rho} &\defeq& \begin{stackTL}
(t_1:\rho_1\tau,t_2:\rho_2\tau).\\
\quad(\direction = \mathcd{true}  \limp (t_1,t_2)\in\Vrel{\tau}{\rho}) ~\land~\\
\quad(\direction = \mathcd{false}  \limp (t_2,t_1)\in\Vrel{\tau}{\reverse{\rho}}) \\
\end{stackTL} \\
\Crelsym{\tau}{\rho} &\defeq& \begin{stackTL}
(t_1:\rho_1\tau,t_2:\rho_2\tau).\\
\quad(\direction = \mathcd{true}  \limp (t_1,t_2)\in\Crel{\tau}{\rho}) ~\land~\\
\quad(\direction = \mathcd{false}  \limp (t_2,t_1)\in\Crel{\tau}{\reverse{\rho}}) \\
\end{stackTL} \\
e_1 \ciuleqo e_2 &\defeq& \begin{stackTL}
(\direction = \etrue \limp e_2 \stepmany e_1) ~\land~\\
(\direction = \efalse \limp e_1 \ciuleq e_2)
\end{stackTL} \\
e_1 \ciuleqt e_2 &\defeq& \begin{stackTL}
(\direction = \etrue \limp e_1 \ciuleq e_2) ~\land~\\
(\direction = \efalse \limp e_2 \stepmany e_1)
\end{stackTL} \\
\end{array}
\]
\mycaption{Symmetric Version of the $\fmu$ Logical Relation and Related Definitions}
\label{fig:symmetric-defs}
\end{figure}

Thus, instead, we define the symmetric relation as shown in
Figure~\ref{fig:symmetric-defs}.  Here, $d$ is a value variable of
type $\tbool$ that we assume is bound in the context in which
these symmetric relations appear.  When $\direction$ is \cd{true},
$\Crelsym{\tau}{\rho}$ and $\Vrelsym{\tau}{\rho}$ are equivalent to
the asymmetric logical relation in one direction; and when
$\direction$ is \cd{false}, they are equivalent to the asymmetric
relation in the other direction.
Thus, by proving two terms to be symmetrically-related in a context
where $\direction$'s identity is unknown, we can effectively prove
logical approximation in both directions simultaneously.

This formulation has several nice properties.  First, it is
straightforward to show that if we take each case of the definition of
$\Vrel{\tau}{\rho}$ in Figure~\ref{fig:logrel}, replace all
occurrences of $\Vrel{\tau}{\rho}$ and $\Crel{\tau}{\rho}$ with their
symmetric versions, and substitute $\equiv$ for $\defeq$, we have a
set of valid relational equivalences.  The same goes for the
relational equivalences in Proposition~\ref{prop:logrel:typesubst}.  (The
same is not true, however, for the definition of $\Crel{\tau}{\rho}$,
because it is inherently asymmetric.)  

The proofs of these symmetric relational equivalences are all
quite easy---each one splits into two cases,
one for $\direction = \etrue$ and one for $\direction = \efalse$.
Here, we sketch the proof for the recursive type case, which is
the most interesting since it uses the \rul{l\"ob} rule.

\begin{prop}
\label{prop:symrec} 
$\Vrelsym{\trec{\alpha}{\tau}}\rho \equiv   \begin{stackTL}
    \mu r.(x_1\valof\rho_1(\trec{\alpha}{\tau}),x_2\valof\rho_2(\trec{\alpha}{\tau})).\\
    \quad \exists y_1,y_2.\ x_1 = \eroll{y_1} \ \land\  x_2 = \eroll{y_2} \ \land \ \\ \quad \quad
     \lift(y_1,y_2) \in \Vrelsym{\tau}\rho,\alpha\mapsto(\rho_1(\trec{\alpha}{\tau}),\rho_2(\trec{\alpha}{\tau}),r)
  \end{stackTL}
$
\end{prop}

\proof Let $R_1$ and $R_2$ denote the relations on the left and right
sides of the equivalence, respectively.  By the \rul{l\"ob} rule, we can
assume that $\later(R_1 \equiv R_2)$.  By Canonical Forms, either
$\direction=\etrue$ or $\direction=\efalse$:
\begin{cases}
  \casen{$\direction=\etrue$} \zilch\\
Unrolling definitions, the proof reduces
  to showing that
  $\later(y_1,y_2)\in\Vrel{\tau}\rho,\alpha\mapsto(\ldots,R_1)$
  iff
  $\later(y_1,y_2)\in\Vrel{\tau}\rho,\alpha\mapsto(\ldots,R_2)$.
This follows from the basic axioms together with the \rul{l\"ob}-inductive
hypothesis $\later(R_1 \equiv R_2)$.

\casen{$\direction=\efalse$} \zilch\\
Similarly, the proof reduces to showing that
  $\later(y_2,y_1)\in\Vrel{\tau}\reverse{\rho},\alpha\mapsto(\ldots,\reverse{R_1})$ iff
  $\later(y_2,y_1)\in\Vrel{\tau}\reverse{\rho},\alpha\mapsto(\ldots,\reverse{R_2})$.
Again, this follows from the basic axioms together with the \rul{l\"ob}-inductive
hypothesis $\later(R_1\equiv R_2)$.
\qed
\end{cases}

\begin{figure*}[!t]
\begin{myfig}
\small
\[
\inferno{sym-exp}{
\ttwf{\con}{\erelatedsym{e_1}{e_2}{\tau}{\rho}}
}{
\begin{array}{c}
\con \ts e_1 : \rho_1\tau \quad
\con \ts e_2 : \rho_2\tau \\[\bp]
\ttwf{\con}{e_1 \stepmany e_1'} \quad
\ttwf{\con}{e_2 \stepmany e_2'} \quad
\ttwf{\con}{\erelatedsym{e_1'}{e_2'}{\tau}{\rho}}
\end{array}
}
\]
\betweenrules
\[
\inferno{sym-exp-$\later$}{
\ttwf{\con}{\erelatedsym{e_1}{e_2}{\tau}{\rho}}
}{
\begin{array}{c}
\con \ts e_1 : \rho_1\tau \quad
\con \ts e_2 : \rho_2\tau \\[\bp]
\ttwf{\con}{e_1 \stepone e_1'} \quad
\ttwf{\con}{e_2 \stepone e_2'} \quad
\ttwf{\con}{\later\erelatedsym{e_1'}{e_2'}{\tau}{\rho}}
\end{array}
}
\]
\betweenrules
\[
\inferno{sym-red}{
\ttwf{\con}{\erelatedsym{e_1}{e_2}{\tau}{\rho}}
}{
\ttwf{\con}{e_1' \stepzero e_1} \quad
\ttwf{\con}{e_2' \stepzero e_2} \quad
\ttwf{\con}{\erelatedsym{e_1'}{e_2'}{\tau}{\rho}}
}
\]
\betweenrules
\[
\inferno{sym-ciu}{
\ttwf{\con}{\erelatedsym{e_1}{e_2}{\tau}{\rho}}
}{
\begin{array}{c}
\con \ts e_1 : \rho_1\tau \quad
\con \ts e_2 : \rho_2\tau \\[\bp]
\ttwf{\con}{\erelatedsym{e_1'}{e_2'}{\tau}{\rho}} \quad
\con \ts e_1' \ciuleqo e_1 \quad
\con \ts e_2' \ciuleqt e_2
\end{array}
}
\]
\betweenrules
\[
\inferno{sym-bind}{
\ttwf{\con}{\erelatedsym{E_1[e_1]}{E_2[e_2]}{\tau'}{\rho'}}
}{
\begin{array}{c}
\con \ts E_1 : \rho_1\tau \eto \rho'_1\tau' \quad
\con \ts E_2 : \rho_2\tau \eto \rho_2'\tau' \quad
\ttwf{\con}{\erelatedsym{e_1}{e_2}{\tau}{\rho}} \\[\bp]
\ttwf{\con,x_1,x_2,\vrelatedsym{x_1}{x_2}{\tau}{\rho},x_1 \ciuleqo e_1, x_2 \ciuleqt e_2}{\erelatedsym{E_1[x_1]}{E_2[x_2]}{\tau'}{\rho'}}
\end{array}
}
\]
\mycaption{Symmetric Versions of Several Derivable Rules}
\label{fig:symmetric}
\end{myfig}
\end{figure*}

\noindent Furthermore, we can easily derive symmetric versions of most of our
derived rules.  In most cases, including all the compatibility
properties, the symmetric rule looks like the asymmetric one, except
with $\mathcal{E}^\approx$ and $\mathcal{V}^\approx$ in place of
$\mathcal{E}$ and $\mathcal{V}$.  Exceptions to this pattern include
the rules from \rul{exp} to \rul{bind2} in Figure~\ref{fig:derivable}.
In Figure~\ref{fig:symmetric}, we give symmetric versions of several
of these, the last two of which employ the relations $e_1 \ciuleqo
e_2$ and $e_1 \ciuleqt e_2$ defined in
Figure~\ref{fig:symmetric-defs}.  These relations are merely a
technical device to enable a symmetric presentation of certain
premises that have the form $e_2 \evalsto e_1$ for one direction of
approximation and $e_1 \ciuleq e_2$ for the other direction.  The
proofs of these rules are all completely straightforward, relying
heavily on the fact (from Proposition~\ref{prop:ciu-props}) that $e_1
\evalsto e_2$ implies $e_1 \ciueqv e_2$.  (Note that the context
$\con$ appearing in all these rules is assumed to bind $\direction$ in
its variable context and contain $\direction\valof\tbool$ in its
proposition context.)

To give the reader a concrete sense of how these rules work,
we present in the next section three detailed examples of how to use
them to prove contextual equivalences.

Finally, since LSLR is inspired by Plotkin and Abadi's logic for
parametricity, one might expect to see some rule corresponding to
``identity extension.''
Denoting contextual equivalence at type $\sigma$ by $\ctxeqv_\sigma$, 
identity extension would mean that, for any open type $\alpha \vdash
\tau$, we would have that $\Crelsym{\tau}{(\alpha\mapsto\ctxeqv_\sigma)}$
equals $\ctxeqv_{\tau[\sigma/\alpha]}$. 
In fact, we do not have such a rule since, as we discovered in the
course of carrying out this work, identity extension does not hold for
the step-indexed model!
For identity extension to hold, one would need that contextual
equivalence at any $\tau$ should \emph{equal} the semantics of
$\Crelsym{\tau}{}$, but it only equals the subset of
$\Crelsym{\tau}{}$ for which the relation holds for all $n$, \ie
roughly, the subset $\{(e_1,e_2) \,\mid\, \forall n. \interp{(e_1,e_2) \in
\Crelsym{\tau}{}}{n}{}\}$.  
The identity extension lemma has traditionally been used to
prove representation independence results, aka 
\emph{free theorems}~\cite{Wadler:89}, and, for pure calculi,
definability results for types~\cite{plotkin-abadi-93}.
In spite of the lack of identity extension we are still able
to prove some free theorems, as we demonstrate in
Section~\ref{sec:free-theorem}.

%%% Local Variables: 
%%% mode: latex
%%% TeX-master: "main"
%%% End: 

%% file: examples.tex
\section{Examples}
\label{sec:examples}

We now show three examples of how to use our \thelogic-based logical
relation to prove interesting contextual equivalences.

The first example is from Crary and Harper~\cite{crary-harper-2007}
(who adapted it from one in Sumii and Pierce~\cite{sumii-pierce-jacm})
and concerns representation independence of ``objects'' with
existential recursive type.  The second, from Sumii and
Pierce~\cite{sumii-pierce-jacm}, is concerned with proving the
syntactic minimal invariant property associated with a general
recursive type~\cite{pitts-reldom,birkedal,crary-harper-2007}.
The third is a canonical example of a Wadler-style ``free
theorem''~\cite{Wadler:89}.

We reason informally in \thelogic{} but present the proofs in some
detail to emphasize the use of the derivable rules from
Section~\ref{sec:logrel}.  Observe that the proofs do not involve any
mention of step indices!

\newcommand{\enot}{\tmfont{not}}
\newcommand{\eeven}{\tmfont{even}}
\newcommand{\tfld}[1]{\tyfont{fld}_{#1}}
\newcommand{\tflag}{\tyfont{flag}}
\newcommand{\ebflag}{\tmfont{bflag}}
\newcommand{\eiflag}{\tmfont{iflag}}
\newcommand{\ebflip}{\tmfont{bflip}}
\newcommand{\eiflip}{\tmfont{iflip}}
\newcommand{\ebret}{\tmfont{bret}}
\newcommand{\eiret}{\tmfont{iret}}

\subsection{Flag Objects} 
\label{sec:flag-objects}
Consider the following type for flag objects,
which have an instance variable (with abstract type $\alpha$) and two
methods.  The first method returns a new object whose flag is 
reversed, while the second method returns the current state of the 
flag.  
\predisp
\[
\begin{array}{lcl}
\tfld{\alpha} &=& 
\trec{\beta}{\tpair{\alpha}
                   {(\tpair{(\tfun{\beta}{\beta})}
                           {(\tfun{\beta}{\tbool})})}} \\
\tflag &=& \texist{\alpha}{\tfld{\alpha}}
\end{array}
\postdisp
\]
We consider two implementations of flags, in which the hidden flag
state is represented by a $\tbool$ and an $\tint$, respectively.  We
assume that $\enot{\,:\,}\tfun{\tbool\!}{\!\tbool}$ and
$\eeven{\,:\,}\tfun{\tint\!}{\!\tbool}$ are implemented in the obvious 
way.   
\predisp
\[
\begin{array}{@{}r@{~}c@{~}l}
\ebflag &=& 
  \epack{\tbool}
        {(\efold{\epair{\etrue}{\epair{\ebflip}{\ebret}}})}
        {\tflag} \\
\ebflip &=& 
  \efun{x}{\tfld{\tbool}}
       {\efold{\langle \begin{stackTL}
                \eapp{\enot}{(\efst{(\eunfold{x})})}, 
                {\esnd{(\eunfold{x})}} \rangle \end{stackTL}}} \\
\ebret &=&
  \efun{x}{\tfld{\tbool}}
       {\efst{(\eunfold{x})}} 
\\[7pt]
\eiflag &=& 
  \epack{\tint}
        {(\efold{\epair{0}{\epair{\eiflip}{\eiret}}})}
        {\tflag} \\
\eiflip &=& 
  \efun{x}{\tfld{\tint}}
       {\efold{\langle \begin{stackTL}
                1+(\efst{(\eunfold{x})}), 
                {\esnd{(\eunfold{x})}} \rangle \end{stackTL}}} \\
\eiret &=&
  \efun{x}{\tfld{\tint}}
       {\eapp{\eeven}{(\efst{(\eunfold{x})})}} \\
\end{array}
\postdisp
\]
To prove equivalence of $\ebflag$ and $\eiflag$, it suffices to show
$\direction,\direction\valof\tbool \ts (\ebflag,\eiflag) \in \Crelsym{\tflag}$.
Equivalently, by Rule~\rul{val}, since both terms are values, it is
enough to show that $\direction,\direction\valof \tbool \ts (\ebflag,\eiflag) \in
\Vrelsym{\tflag}$.  Unfolding the definition of
$\Vrelsym{\texist{\alpha}{\tfld{\alpha}}}$, we choose 
$\alpha_1{\,\mapsto\,}\tbool$, $\alpha_2{\,\mapsto\,}\tint$,
$y_1{\,\mapsto\,}v_1$, $y_2{\,\mapsto\,}v_2$, and $r{\,\mapsto\,}R$
as the substitution for its existentially-bound variables, where
$v_1 = \efold{\epair{\etrue}{\epair{\ebflip}{\ebret}}}, v_2 = \efold{\epair{0}{\epair{\eiflip}{\eiret}}}$, and
\predisp
\[
\begin{array}{@{}r@{~}c@{~}l}
R &=& \begin{stackTL}(x_1 \valof \tbool,x_2 \valof \tint).\,
       \exists y \valof \tint.\,
         (x_1=\etrue \land 2y \termto x_2)
%         \qquad\qquad\qquad\qquad~\lor~
\lor         (x_1=\efalse \land 2y{\,+}1 \termto x_2)
      \end{stackTL} 
\end{array}
\postdisp
\]
Let $\rho=\alpha{\,\mapsto\,}(\tbool,\tint,R)$.  It now suffices to
show $(v_1,v_2)\in\Vrelsym{\tfld{\alpha}}\!\rho$, or equivalently
(using the compatibility rules and several applications of Rule~\rul{val}):
\begin{enumerate}[(1)]
\item Show $(\etrue,0) \in \Vrelsym{\alpha}\rho$. This is 
  immediate from the definition of $R$ by choosing $y\mapsto 0$. 
\item Show
  $(\ebflip,\eiflip)\in\Vrelsym{\tfun{\tfld{\alpha}}{\tfld{\alpha}}}\rho$.
  By the compatibility rule for functions, we assume that
  $(x_1, x_2) \in$ $\Vrelsym{\tfld{\alpha}}\rho$, and are required to
  show: 
\predisp
\[
\begin{array}{l}
(\begin{stackTL}
 \efold{\epair{\eapp{\enot}{(\efst{(\eunfold{x_1})})}} 
               {\esnd{(\eunfold{x_1})}}}, \\
 \efold{\epair{1+(\efst{(\eunfold{x_2})})}
              {\esnd{(\eunfold{x_2})}}}) 
\end{stackTL}\\
\qquad \in \Crelsym{\tfld{\alpha}}\rho
\end{array}
\postdisp
\]
By compatibility, we have that
$(\efst{(\eunfold{x_1\!})},\efst{(\eunfold{x_2})}){\in\,}\Crelsym{\alpha}{\!\rho}$. Thus,
by Rule~\rul{sym-bind}, we can assume that $(z_1,z_2) \in \Vrelsym{\alpha}{\rho} \equiv
R$ for some $z_1$ and $z_2$, and the proof reduces to showing \predisp
\[
\begin{array}{l}
(\begin{stackTL}
 \efold{\epair{\eapp{\enot}{z_1}}
               {\esnd{(\eunfold{x_1})}}}, \\
 \efold{\epair{1+z_2}
              {\esnd{(\eunfold{x_2})}}}) \in \Crelsym{\tfld{\alpha}}\rho
\end{stackTL}\\
\end{array}
\postdisp
\]
By compatibility again, this reduces to
showing that $(\eapp{\enot}{z_1},1+z_2)\in \Crelsym{\alpha}{\rho}$.
By Rule~\rul{sym-exp}, it simply remains to show that $\eapp{\enot}{z_1}$
and $1+z_2$ evaluate to values that are related by $R$.  The following
lemma suffices:
\predisp
\[
\begin{array}{l}
\forall z_1,z_2.\,(z_1,z_2)\in R \limp 
\exists z_1',z_2'.\ \eapp{\enot}{z_1}\termto z_1' \land 1+z_2 \termto z_2'
\land (z_1',z_2')\in R
\end{array}
\postdisp
\]
Expanding out the definition of membership in $R$, we arrive at a
strictly first-order statement that is provable by straightforward
means in the meta-logic.

\item Show
  $(\ebret,\eiret)\in\Vrelsym{\tfun{\tfld{\alpha}\!}{\tbool}}\rho$.
  This is similar to part~(2), with the proof boiling down to the
  first-order statement
\predisp
\[
\begin{array}{l}
\forall z_1,z_2.\,(z_1,z_2)\in R \limp 
\eapp{\tmfont{even}}{z_2} \termto z_1
\end{array}
\postdisp\eqno{\qEd}
\]
\end{enumerate}

\newcommand{\eid}{\tmfont{id}}

\subsection{Syntactic Minimal Invariance} 
\label{sec:syntactic-projection}
\label{sec:minimal-invariance}

The proof of our next example relies on Canonical Forms, a first-order
lemma about $\fmu$ that we assume is proven outside LSLR by
traditional means.  This standard lemma, which characterizes the shape
of well-typed values, is only available to us because (following
Pitts~\cite{pitts:attapl}) we have constructed the logical relation
from syntactically well-typed terms.  For further discussion of this
point, see Section~\ref{sec:lics-comparison}.

Let $\tau = \trec{\alpha}{\tsum{\tunit}{(\tfun{\alpha}{\alpha})}}$.
We are going to show that the identity function $\eid =
\efun{x}{\tau}{x}$ is equivalent to \predisp
\[
\begin{array}{@{}r@{~}c@{~}l}
v & = & 
\efix{f}{x{\,:\,}\tau}
{\ecasebrneg{(\eunfold{x})}{\_}{\efold{(\einl{\eunit})}}
                    {g}{\efold{(\einr{(\efun{y}{\tau}{f(g(f\,y))})})}}}
\end{array}
\postdisp
\]
This corresponds to the \emph{minimal invariant} property in the
domain-theoretic work of Pitts \cite{pitts-reldom}, which Birkedal and
Harper subsequently proved in an operational setting~\cite{birkedal}.

To prove contextual equivalence of $\eid$ and $v$, we can show
$\direction,\direction\valof\tbool\ts\,(\eid,v)\in\Vrelsym{\tfun{\tau}{\tau}}$.
Our proof will be parametric in $d$.  By the \rul{l\"ob} rule, we assume
$\later(\eid,v)\in\Vrelsym{\tfun{\tau}{\tau}}$ and proceed to prove
$(\eid,v)\in\Vrelsym{\tfun{\tau}{\tau}}$.  Now, by
(the symmetric version of) Rule~\rul{funext}
and~\rul{sym-exp}, we assume $(x_1,x_2)\in\Vrelsym{\tau}$, and it suffices
to show \predisp
\[
\begin{array}{@{}l}
(x_1,{\ecasebrneg{(\eunfold{x_2})}{\_}{\efold{(\einl{\eunit})}}
                    {g}{\efold{(\einr{(\efun{y}{\tau}{v(g(v\,y))})}))\in \Crelsym{\tau}}}}
\end{array}
\postdisp
\]
By relatedness of $x_1$ and $x_2$, we know that there exist
$y_1$ and $y_2$ such that $x_1 = \efold{y_1}$, $x_2 = \efold{y_2}$,
and $\later(y_1,y_2)\in\Vrelsym{\tsum{\tunit}{(\tfun{\tau}{\tau})}}$.
By Canonical Forms, since $y_2 \valof
\tsum{\tunit}{(\tfun{\tau}{\tau})}$, we know that either $y_2 =
\einl{\eunit}$ or there exists $y_2'$ such that $y_2 = \einr{y_2'}$.
In either case, there exists $z \valof \tsum{\tunit}{(\tfun{\tau}{\tau})}$
such that the $\tmfont{case}$ expression above evaluates to
$\efold{z}$.  Consequently, by Rule~\rul{sym-exp},
the goal reduces to showing
\[
(\efold{y_1},\efold{z}) \in \Vrelsym{\trec{\alpha}{\tsum{\tunit}{(\tfun{\alpha}{\alpha})}}}
\]
Unfolding the definition of $\Vrelsym{\trec{\alpha}{\tsum{\tunit}{(\tfun{\alpha}{\alpha})}}}$, it suffices by Rule~\rul{weak-$\later$} to show
\[
(y_1,z) \in \Vrelsym{\tsum{\tunit}{(\tfun{\tau}{\tau})}}
\]
under a strengthened (\ie $\earlier$'d) context in which the $\later$
has been removed from any of our previous assumptions.  In particular,
we may now assume our \rul{l\"ob}-inductive hypothesis
$(\eid,v)\in\Vrelsym{\tfun{\tau}{\tau}}$, as well as 
$(y_1,y_2)\in\Vrelsym{\tsum{\tunit}{(\tfun{\tau}{\tau})}}$, to hold
``now'' as opposed to ``later''.  The latter assumption yields two
cases:

\begin{cases}
\casen{$\tmfont{inl}$} \zilch\\
$y_1=y_2=z=\einl{\eunit}$.  Trivial.

\casen{$\tmfont{inr}$} \zilch\\
$y_1=\einr{y_1'}$,
$y_2=\einr{y_2'}$, $(y_1',y_2')\in\Vrelsym{\tfun{\tau}{\tau}}$, and
$z = \einr{(\efun{y}{\tau}{v\,(y_2'\,(v\,y))})}$.
Thus, to complete the proof it suffices to show
\predisp
\[
\begin{array}{@{}l}
(y_1',\efun{y}{\tau}{v\,(y_2'\,(v\,y))}) \in \Vrelsym{\tfun{\tau}{\tau}} 
\end{array}
\postdisp
\]
Applying Rule~\rul{funext} (in its symmetric form) and
Rule~\rul{sym-exp}, we assume $(z_1,z_2)\in\Vrelsym{\tau}$, and
have to show
\predisp
\[
\begin{array}{@{}l}
(y_1'\,z_1,v\,(y_2'\,(v\,z_2)))\in\Crelsym{\tau}
\end{array}
\postdisp
\]
From $(\eid,v)\in\Vrelsym{\tfun{\tau}{\tau}}$, together with relatedness of $z_1$ and $z_2$, we may conclude by Rules~\rul{app} and~\rul{sym-red}
that $(z_1,v\,z_2)\in\Crelsym{\tau}$.  By relatedness of $y_1'$ and $y_2'$
and Rule~\rul{app}, we have that $(y_1'\,z_1,(y_2'\,(v\,z_2)))\in\Crelsym{\tau}$.  Thus, by Rule~\rul{sym-bind}, choosing as the evaluation
contexts of interest $\hole$ and $v\,\hole$, our goal reduces to
showing that for any $z_1',z_2'$, if $(z_1',z_2')\in\Vrelsym{\tau}$, then $(z_1',v\,z_2')\in\Crelsym{\tau}$.
As before, this follows from $(\eid,v)\in\Vrelsym{\tfun{\tau}{\tau}}$,
together with Rules~\rul{app} and~\rul{sym-red}.
\qed
\end{cases}

\subsection{A ``Free Theorem''}
\label{sec:free-theorem}

\newcommand{\epolyid}{\tmfont{Id}}

Suppose that $\tau$ and $\sigma$ are closed types, that $h$ and $f$
are values such that $h :
\tall{\alpha}{\tfun{\alpha}{\tfun{\alpha}{\alpha}}}$ and $f :
\tfun{\tau}{\sigma}$, and that $v$ and $w$ are values of type $\tau$.
We will prove that $\etapp{h}{\sigma}\,(f\,v)\,(f\,w)$ contextually
approximates $f\,(\etapp{h}{\tau}\,v\,w)$ unconditionally, and that
the reverse approximation also holds if $f$ is total (a sufficient, 
but not necessary, condition), defined as $\mathsf{total}(f) \defeq
\forall x.\ x\valof\tau \limp \exists y.\ f\,x \termto y$.

The proof is interesting in that it is mostly done in a symmetric
fashion, except for one inner lemma, which requires us to split into
cases, one for each asymmetric direction of approximation.  Since
one of the two directions includes an extra assumption concerning the
totality of $f$, we will actually prove the theorem
\[
\con \ts (\etapp{h}{\sigma}\,(f\,v)\,(f\,w),~f\,(\etapp{h}{\tau}\,v\,w))
\in \Crelsym{\sigma}
\]
where $\con = d,d \valof \tbool, d = \efalse \limp \mathsf{total}(f)$.
To prove the theorem, we use Rule~\rul{sym-bind} with the evaluation
contexts $\hole$ and $f\,\hole$, respectively.  The proof is in two parts.  

\vspace{1ex}
\noindent
\textbf{Part 1~}
First, we prove that
\predisp
\[
\begin{array}{@{}l}
(\etapp{h}{\sigma}\,(f\,v)\,(f\,w), \etapp{h}{\tau}\,v\,w) \in 
\Crelsym{\alpha}\rho
\end{array}
\postdisp
\]
where $\rho = \alpha \mapsto (\sigma,\tau,R)$ and 
\predisp
\[
\begin{array}{@{}l}
R = (y_1\valof\sigma,y_2\valof\tau).\ (y_1,f\,y_2)\in \Crelsym{\sigma}
\postdisp
\end{array}
\]
By Theorem~\ref{thm:parametricity},
$(h,h)\in\Crelsym{\tall{\alpha}{\tfun{\alpha}{\tfun{\alpha}{\alpha}}}}$.
Thus,
$(h\,\sigma,h\,\tau)\in\Crelsym{\tfun{\alpha}{\tfun{\alpha}{\alpha}}}\rho$.
To prove our desired result (by Rule~\rul{app}), it remains to
show that $(f\,v,v)\in\Crelsym{\alpha}{\rho}$ and
$(f\,w,w)\in\Crelsym{\alpha}{\rho}$.  We show the proof for the former;
the latter is exactly the same.

This is the place where we need to split into cases depending on the
direction of the proof.  Both cases use the fact, due to the
Fundamental Theorem, that $(f\,v,f\,v)\in\Crelsym{\sigma}$.

\begin{cases}
  \casen{$\direction = \etrue$} \zilch\\
  We need to show $(f\,v,v) \in \Crel{\alpha}\rho$.  Since
  $(f\,v,f\,v)\in\Crel{\sigma}$, by Rule~\rul{bind} (using evaluation
  context $\hole$) and Rule~\rul{val} we may assume that there exist
  $x_1,x_2$ such that $(x_1,x_2) \in \Vrel{\sigma}$ and $x_2 \ciuleq
  f\,v$, and it remains to show $(x_1,v)\in\Vrel{\alpha}\rho = R$.
  The latter is equivalent to $(x_1,f\,v)\in\Crel{\sigma}$, which
  follows directly from the assumptions by Rule~\rul{ciu}.

  \casen{$\direction = \efalse$} \zilch\\
  We need to show $(v,f\,v) \in \Crel{\alpha}{\reverse\rho}$.  Using
  the assumption $\mathsf{total}(f)$, which is available since
  $\direction = \efalse$, we know that there exists $x\valof\sigma$
  such that $f\,v \termto x$.  Thus, $f\,v \ciuleq x$ and $x \ciuleq
  f\,v$.  By Rules~\rul{ciu} and~\rul{val}, it suffices to show
  $(v,x)\in\Vrel{\alpha}{\reverse\rho} = \reverse{R}$.  Unrolling the
  definition of $R$, we see that the goal is equivalent to
  $(f\,v,x)\in\Crel{\sigma}$, which follows from
  $(f\,v,f\,v)\in\Crel{\sigma}$ and $f\,v\ciuleq x$ by
  Rule~\rul{ciu}.
\end{cases}

\vspace{1ex}
\noindent
\textbf{Part 2~}
Next, we assume that $(z_1,z_2)
\in \Vrelsym{\alpha}\rho \equiv R$ and we need to
show that 
\[
\begin{array}{@{}l}
(z_1,f\,z_2) \in \Crelsym{\sigma}
\postdisp
\end{array}
\]
But this falls out directly from the definition of $R$, so we are done.
\qed

%%% Local Variables: 
%%% mode: latex
%%% TeX-master: "main"
%%% End: 

%% file: discussion.tex
\section{The Merits of Our Approach}
\label{sec:discussion}

By way of comparison with previous work, we now informally present an
alternative proof of the ``flag objects'' example (from
Section~\ref{sec:flag-objects}) in the style of
Ahmed~\cite{ahmed-2006}.  Following that, we discuss how our LSLR
proof relates to and improves on this alternative proof.

\subsection{Flag Objects Proof With Explicit Step Manipulation}
We now sketch a proof for the ``flag objects'' example from
Section~\ref{sec:flag-objects} using Ahmed's logical
relation~\cite{ahmed-2006}.  Since the latter is asymmetric, to prove
equivalence of $\ebflag$ and $\eiflag$ at type $\tflag$, we must show
that for all $n \geq 0$, $(n, \ebflag, \eiflag) \in \Crel{\tflag}$ and
$(n,\eiflag, \ebflag) \in \Crel{\tflag}$, where $\Crel{\cdot}$ is the
asymmetric logical relation for closed terms from Ahmed's paper.
Here, writing $(n,e_1,e_2) \in \Crel{\tau}$ means that $e_1$ and $e_2$
are related for $n$ steps---or more specifically, that if $e_1$
terminates in less than $n$ steps then $e_2$ will terminate (in any
number of steps) and the resulting values will be related for the
remaining number of steps.  We discuss only one direction of the
proof; the other direction is similar.

To prove that $(n,\ebflag,\eiflag) \in \Crel{\tflag}$ for arbitrary
$n\geq 0$, it suffices to show $(n,\ebflag,\eiflag) \in \Vrel{\tflag}$,
since $\ebflag$ and $\eiflag$ are values.  We take $\tau_1 = \tbool$,
$\tau_2 = \tint$, and \predisp
\[
\begin{array}{@{}r@{~}c@{~}l}
R &=& \{\,(n',v_1,v_2) \mid~\begin{stackTL}\ts v_1 : \tbool ~\land~
  \ts v_2 : \tint ~\land~ \\
\exists y : \tint.\,
         (v_1=\etrue \land 2y \termto v_2)
\lor         (v_1=\efalse \land 2y{\,+}1 \termto v_2)\}
      \end{stackTL} 
\end{array}
\postdisp
\]
Let $\rho=\alpha{\,\mapsto\,}(\tbool,\tint,R)$.  It then suffices to
show, for all $m<n$, that
\predisp
\[
\begin{array}{@{}l}
(m,\efold{\epair{\etrue}{\epair{\ebflip}{\ebret}}},\efold{\epair{0}{\epair{\eiflip}{\eiret}}})
\in \Vrel{\tfld{\alpha}}\rho
\end{array}
\postdisp
\]
Unwinding the definitions of $\Vrel{\trec{\beta}{\tau}}$ and
$\Vrel{\tpair{\tau_1}{\tau_2}}$, it now suffices to show the following
for all $k<m$: 
\begin{enumerate}[(1)]
\item Show $(k,\etrue,0) \in \Vrel{\alpha}\rho$.  This is immediate
  from the definition of $R$, choosing $y = 0$ as before. 
\item Show $(k,\ebflip,\eiflip) \in
  \Vrel{\tfun{\tfld{\alpha}}{\tfld{\alpha}}}\rho$.  For arbitrary
  $j<k$, assuming we are given $(j,v_{a1},v_{a2}) \in
  \Vrel{\tfld{\alpha}}\rho$, we are required to show: \predisp
\[
\begin{array}{l}
(j,\begin{stackTL}
   \efold{\epair{\eapp{\enot}{(\efst{(\eunfold{v_{a1}})})}} 
               {\esnd{(\eunfold{v_{a1}})}}}, \\
   \efold{\epair{1+(\efst{(\eunfold{v_{a2}})})}
               {\esnd{(\eunfold{v_{a2}})}}}) 
\in \Crel{\tfld{\alpha}}\rho
\end{stackTL}
\end{array}
\postdisp
\]
We assume that
$\efold{\epair{\eapp{\enot}{(\efst{(\eunfold{v_{a1}})})}}{\esnd{(\eunfold{v_{a1}})}}}$
evaluates to a value $v_{f1}$ in $i<j$ steps.  We are required to show
that there exists a value $v_{f2}$ such that
$\efold{\epair{1+(\efst{(\eunfold{v_{a2}})})}{\esnd{(\eunfold{v_{a2}})}}}$
evaluates to $v_{f2}$ and $(j-i,v_{f1},v_{f2}) \in \Vrel{\tfld{\alpha}}\rho$. 
Since these expressions clearly require more than one step of evaluation,
we know that $j > 2$ (which is relevant here when we talk about $j-1$).

From $(j,v_{a1},v_{a2}) \in \Vrel{\tfld{\alpha}}\rho$, it follows that
$v_{a1} = \efold{v_{10}}$ and $v_{a2} = \efold{v_{20}}$, and
furthermore that $v_{10} = \epair{v_{11}}{v_{12}}$ and $v_{20} =
\epair{v_{21}}{v_{22}}$, where $(j-1,v_{11},v_{21}) \in
\Vrel{\alpha}\rho$ and $(j-1,v_{12},v_{22}) \in
\Vrel{\tpair{(\tfun{\tfld{\alpha}}{\tfld{\alpha}})}{(\tfun{\tfld{\alpha}}{\tbool})}}\rho$. 

Hence, by the operational semantics, we have that: 
\predisp
\[
\begin{array}{@{}l}
\efold{\epair{\eapp{\enot}{(\efst{(\eunfold{v_{a1}})})}}{\esnd{(\eunfold{v_{a1}})}}}
\step\\
\efold{\epair{\eapp{\enot}{(\efst{v_{10}})}}{\esnd{(\eunfold{v_{a1}})}}}
\step\\
\efold{\epair{\eapp{\enot}{v_{11}}}{\esnd{(\eunfold{v_{a1}})}}}
\step\\
\efold{\epair{\neg{v_{11}}}{\esnd{(\eunfold{v_{a1}})}}}
\step\\
\efold{\epair{\neg{v_{11}}}{\esnd{v_{10}}}}
\step\\
\efold{\epair{\neg{v_{11}}}{v_{12}}}\\
= v_{f1}
\end{array}
\postdisp
\]
where $\neg{v_{11}}$ is a value denoting the negation of $v_{11}$. 

Also, by the operational semantics: 
\predisp
\[
\begin{array}{@{}l}
\efold{\epair{1+(\efst{(\eunfold{v_{a2}})})}{\esnd{(\eunfold{v_{a2}})}}}
\step\\
\efold{\epair{1+(\efst{v_{20}})}{\esnd{(\eunfold{v_{a2}})}}}
\step\\
\efold{\epair{1+v_{21}}{\esnd{(\eunfold{v_{a2}})}}}
\step\\
\efold{\epair{1\hat{+}{v_{21}}}{\esnd{(\eunfold{v_{a2}})}}}
\step\\
\efold{\epair{1\hat{+}{v_{21}}}{\esnd{v_{20}}}}
\step\\
\efold{\epair{1\hat{+}{v_{21}}}{v_{22}}}\\
= v_{f2}
\end{array}
\postdisp
\]
where $1\hat{+}{v_{21}}$ is a value denoting the sum of 1 and $v_{21}$. 

It remains for us to show that $(j-i,v_{f1},v_{f2}) \in
\Vrel{\tfld{\alpha}}\rho$.  By the definition of
$\Vrel{\trec{\beta}{\tau}}$ and $\Vrel{\tpair{\tau_1}{\tau_2}}$, it
suffices to show that, assuming $j-i > 0$:
\begin{enumerate}[$\bullet$]
\item $(j-i-1,\neg{v_{11}},1\hat{+}v_{21}) \in \Vrel{\alpha}\rho$, which
  follows from $(j-1,v_{11},v_{21}) \in \Vrel{\alpha}\rho$ and the
  definition of $R$. 
\item $(j-i-1,v_{12},v_{22}) \in 
\Vrel{\tpair{(\tfun{\tfld{\alpha}}{\tfld{\alpha}})}{(\tfun{\tfld{\alpha}}{\tbool})}}\rho$,
which follows from the fact above that $v_{12}$ and $v_{22}$ are related for
$j-1$ steps, which means that they must be related for fewer steps.
\end{enumerate}
\item Show $(k,\ebret,\eiret) \in
  \Vrel{\tfun{\tfld{\alpha}}{\tbool}}\rho$.  This is similar to the
  proof of part (2).  
\qed
\end{enumerate}

%%%%%%%%%%%%%%%%%%%%%%%%%%%%%%%%%%%%%%%%%%%%%%%%%%

\subsection{What Have We Achieved?}

One can see that the above proof requires quite a bit of pedantic step
manipulation that is entirely unimportant in terms of the overall
proof.  The proof using \thelogic{} allows us to ignore steps and
focus on the interesting parts of the proof.

Perhaps more importantly, the above proof is almost ``mindless'' in
the sense that it proceeds by simply unrolling definitions.  For
instance, step (2) of the proof proceeds to prove relatedness of two
terms for $j$ steps in $\Crel{\tfld{\alpha}}\rho$ by symbolically
evaluating them to values and then showing that the resulting values
are related for $j-i$ steps, where $i$ is the number of steps it takes
to evaluate the first term.  This is exactly how one would attempt
to prove the subgoal if one were just to expand the definition
of $\Crel{\tfld{\alpha}}\rho$.  But as a result, one is forced to
talk about the particular number of steps the first term takes to
evaluate, and moreover, the \emph{idea} of the proof is obscured.

In contrast, the LSLR proof of this example has a much clearer
structure because it is constructed using higher-level proof rules.
In the aforementioned step (2), the LSLR proof does not need to
symbolically execute the terms because it is possible to use
compatibility rules, together with the \rul{sym-bind} rule, instead.
This combination is applicable precisely because the two terms being
related have a very similar structure and only differ in one place.
Thus, the ability to prove the relatedness of the terms using those
rules sheds light on \emph{why} they are equivalent.

That said, the reader may wonder: is the logic LSLR really
\emph{necessary}?  Can we take the proof rules that we have derived in
LSLR and interpret them back into the step-indexed model, thus
resulting in proof principles for the step-indexed model that
\emph{do} mention steps but nonetheless help one to write proofs in a
more structured way?  We believe that to some extent this should be
possible.  For example, here is a variant of the \rul{bind2} rule that
holds (ignoring syntactic typing side conditions) for Ahmed's model:
\[
\infer{
(j,E_1[e_1],E_2[e_2]) \in \Crel{\tau'}\rho'
}{
\begin{array}{c}
(j,e_1,e_2) \in \Crel{\tau}\rho
\\
\forall i \leq j.\ \forall v_1,v_2.\ (i,v_1,v_2) \in \Vrel{\tau}\rho \Longrightarrow
(i,E_1[v_1],E_2[v_2]) \in \Crel{\tau'}\rho'
\end{array}
}
\]
This proof principle is almost as clean as the \rul{bind2} rule, the
only difference being that this step-indexed version requires an
explicit quantification over future worlds $i$, whereas in LSLR that
quantification is baked into the interpretation of the logical
judgment.  While this explicit quantification is annoying, the above
rule should still (we believe) be useful in improving the structure of
``direct'' step-indexed proofs.  (It is less clear how to interpret
the symmetric rules from Figure~\ref{fig:symmetric} into useful
step-indexed rules.)

Thus, what we view as the major contribution of this work is the
development of a set of proof principles to enable better structuring
of step-indexed proofs.  By working at the logical level, instead of
directly in the step-indexed model, we have been forced to come up
with clean high-level rules that do not mention steps, but at least
some of these rules should in retrospect also be useful for improving
the structure of direct step-indexed proofs.

%%% Local Variables: 
%%% mode: latex
%%% TeX-master: "main"
%%% End: 

%% file: lics-comparison.tex
\section{Comparison With an Earlier Version of LSLR}
\label{sec:lics-comparison}

In this section, we explain the four main differences between the
present version of LSLR and the earlier version that we described in
our LICS 2009 paper~\cite{dreyer+:lics09}.

\paragraph{\bf Atomic Typing and Value Predicates}

In the earlier version of LSLR, we built in the atomic predicates of
syntactic typing ($e:\tau$) and value-hood ($\OVal$) as primitive
notions in the logic, instead of treating them as ordinary atomic
relations as we do presently.  Specifically, we imposed a distinction
in the variable context $\vcon$ between value variables $x$ and term
variables $t$ and required typing annotations on their context
bindings.  We also required relation variables to be bound with
explicit relation types $\TRel{\tau_1}{\tau_2}$ and
$\VRel{\tau_1}{\tau_2}$ (relations were restricted to be binary).
In the present version, we also make use of relation types, but these
are definable in the logic and need not be made primitive.

There was in retrospect no particularly good reason for giving these
predicates special treatment, nor for restricting the arity of
relations to $2$.  We feel our present treatment is simpler, cleaner,
and more general.

\paragraph{\bf Distinction Between Logic and Model}

In the earlier version of LSLR, we made a distinction between our main
logical judgment, $\con \ts P$, defined by a set of core inference
rules, and its interpretation into the model, which we wrote as $\con
\models P$.  This enabled a more precise characterization of what it
means for a rule (like those in Figure~\ref{fig:derivable}) to
be ``derivable''.

In the present paper, we conflate $\ts$ and $\models$, thus allowing
arbitrary new inference rules to be added to the logic at a later time
as long as they can be proven sound.  We have made this change because
ultimately it is not clear to us why (or that) the core set of
inference rules we gave in Figure~\ref{fig:logic-rules} are the
``right'' (or ``canonical'') ones.  They are simply a set of sound
rules that we have found to be useful for doing nearly all of our
proofs about logical relations in LSLR.  However, as in the LICS
paper, those core rules are not ``complete''---occasionally, as in the
proof of Adequacy of our logical relation, one needs to reason
directly in the model.  We therefore feel there is no particular need
to grant those core rules ``definitional'' status.

\paragraph{\bf Completeness of the Logical Relation}

In the LICS paper, we defined a logical relation for $\fmu$
that---like Ahmed's original logical relation for
$\fmu$~\cite{ahmed-2006}---was sound, but not complete, with respect
to contextual approximation.  (The incompleteness is related to the
treatment of existential types, \cf Example 7.7.4 in
Pitts~\cite{pitts:attapl}.)  The only substantive difference between
that logical relation and our present version is in the definition of
$\Crel{\tau}\rho$.  If $(e_1,e_2)\in\Crel{\tau}\rho$, then in the case
when $e_1\termzero v_1$, the LICS logical relation would insist that
$e_2$ \emph{evaluate to} some value $v_2$ such that
$(v_1,v_2)\in\Vrel{\tau}\rho$.  In our present logical relation, we
only insist that $e_2$ \emph{be ciu-approximated by} such a value
$v_2$.  This added flexibility is important in proving the
Ciu-Transitivity property (Theorem~\ref{thm:ciu-transitivity}), which
is the key to showing completeness of our present version of the
logical relation.

This change to the logical relation has resulted in changes to some of
the derivable rules in Figures~\ref{fig:derivable}
and~\ref{fig:symmetric} as well.  Rule~\rul{ciu}, for instance, is
more flexible than the corresponding Rule~3 in the LICS paper, whereas
Rule~\rul{bind} is more restrictive than the corresponding Rule~6 in
the LICS paper.  Practically speaking, though, these differences seem
to be very minor, and they have not induced any serious changes to our
proofs of the examples in Section~\ref{sec:examples}.

\paragraph{\bf Fixing a Technical Flaw}

Our present account of LSLR fixes a technical flaw in the LICS
version, namely that three inference rules in that paper are unsound
(and all three for similar reasons).  Luckily, none of the rules was
of critical importance.  The common error we made in our proofs for
all three rules was in forgetting that, when reasoning about the
$\later$ operator, the interesting ``base case'' is often not world
$0$ but world $1$.

The first unsound rule is Rule~\rul{$\later\exists1$} from
Figure~\ref{fig:logic-rules}, in the case where $\vcon$ is of the form
$x:\tau$.  (Note: our present version of LSLR does not run afoul of
this bug precisely because we no longer bake typing or value
predicates into the $\vcon$.)  The problem arises when $\tau$ is an
uninhabited type, such as $\forall\alpha.\alpha$.  The
$\later\exists1$ rule says that $\later\exists x:\tau.P$ implies
$\exists x:\tau.\later P$.  In order for this to be sound it must at
least be the case that $\sembrace{\later\exists x:\tau.P}1$ implies
$\sembrace{\exists x:\tau.\later P}1$.  However, the former is
trivially true, and the latter is false because there is no value of
type $\tau$.  The rule is easy to show sound under the side condition
that $\tau$ is inhabited.

The second and third unsound rules are those numbered Rule~10 and
Rule~8 (the backwards direction) in the LICS paper, which are as
follows:
\[
\infer{
\ttwf{\con}{\erelated{\efold{e_1}}{\efold{e_2}}{\trec{\al}{\tau}}{\rho}}
}{
\ttwf{\con}{\later\erelated{e_1}{e_2}{\tau[\trec{\al}{\tau}/\al]}{\rho}}
}
\qquad
\infer{
\ttwf{\con}{\erelated{e_1}{e_2}{\trec{\al}{\tau}}{\rho}}
}{
\ttwf{\con}{\erelated{\eunfold{e_1}}{\eunfold{e_2}}{\tau[\trec{\al}{\tau}/\al]}{\rho}}
}
\]
The problem with these rules, again, is that the implications do not
hold when the propositions are interpreted at world $1$.  In our buggy
proofs of derivability for these rules, the error manifested itself as
a need to derive $e_2 \termto v_2$ in a context where we only knew
$\later(e_2 \termto v_2)$.  Interestingly, $\sembrace{\later(e_2
  \termto v_2)}n$ \emph{does} imply $\sembrace{e_2 \termto v_2}n$ for all
$n$ except $n=1$.

Fortunately, the only one of these rules that we actually made any use
of was the last one.  We used it in the proof of the syntactic minimal
invariance example, and thus our present proof of that example is
somewhat different than the one given in the LICS paper.  In
particular, in proving that example, we now make critical use of the
standard Canonical Forms property for well-typed values, which we did
not in the LICS paper.

%%% Local Variables: 
%%% mode: latex
%%% TeX-master: "main"
%%% End: 

%% file: related.tex
\presec
\section{Related Work and Conclusion}
\label{sec:related}
\label{sec:related-work}
\postsec

As explained in the introduction, \thelogic{} is greatly indebted to
(1) Plotkin and Abadi's logic for parametricity, and (2) Appel,
Melli\`es, Richards, and Vouillon's ``very modal model''. However, there
are also significant differences between our work and theirs.

Plotkin and Abadi's logic was originally developed for pure System F,
as was Abadi, Cardelli and Curien's System
R~\cite{Abadi:Cardelli:Curien:93}.  (The latter is less expressive, in
that the only relations definable in the logic are those that are maps
of System F functions.)  In recent years, several extensions of PAL to
richer languages with effects have been proposed.
Plotkin~\cite{PlotkinGD:secotr} suggested a variant for a second-order
linear type theory with a polymorphic fixed-point combinator to
combine polymorphism with recursion; it relies on an abstract notion
of admissible relations (see also~\cite{BirkedalL:lapl-journal}),
whereas our logic \thelogic{} does not.  Bierman, Pitts and
Russo~\cite{BiermanGM:opeplp} equipped the language suggested by
Plotkin with an operational semantics, resulting in a programming
language called Lily. Here instead we consider a standard
call-by-value language with impredicative polymorphism and recursive
types and show how to define a logic for reasoning about that
language's operational semantics.

The main difference between our work and AMRV's very modal model is
the application: whereas AMRV use the later operator $\later A$ to
reason about type safety (a unary property) in a low-level language,
we use it to reason about contextual approximation and equivalence
(binary properties) in a high-level language.  Certain issues, such as
the development of both symmetric and asymmetric reasoning principles,
do not arise in the unary setting.  There are other concerns that do
not apply to our setting, such as the desire for non-monotone
predicates (hence our monotonicity axiom, which simplifies matters).
Moreover, a significant component of our contribution is the
derivation of a set of useful, language-specific inference rules
and the application of those rules to several representative examples
from the literature.

Our application of the \rul{l\"ob} rule in connection with a
logical-relations method results in coinductive-style reasoning
principles reminiscent of those used in bisimulation-based methods
like Sumii and Pierce's~\cite{sumii-pierce-jacm}, or Lassen and
Levy's~\cite{Lassen-Levy:LICS2008}.  Sumii and Pierce give several
example applications of their method in a language setting very
similar to the one we consider here.  In Section~\ref{sec:examples},
we already showed how to use LSLR to prove two examples adapted from
their paper, and our approach is capable of straightforwardly handling
the other examples that their method can prove as well.

That said, Sumii and Pierce do present one equivalence, the ``IntSet''
example at the beginning of Section 7 of their paper, which does not
seem to be provable \emph{directly} within our logic, although it is
provable through a transitive combination of equivalence proofs.  They
use this example to exhibit a limitation of their method with respect
to reasoning about higher-order functions, and hence to motivate an
``up-to-context'' extension of their bisimulation that alleviates the
problem.  However, they do not actually offer a proof of the IntSet
example (using the up-to-context extension or otherwise), and we
believe the proof to be considerably more involved than for the other
up-to-context examples in their paper.  Ahmed~\cite{ahmed-2006} has
given a proof of this example using step-indexed logical relations
(see her technical report), but her proof is closely tailored to the
specific example and seems difficult to adapt, \eg if the ADT in the
example is extended with a ``remove'' operation.

The IntSet example is challenging because it involves an equivalence
between two recursive functions that are structurally quite dissimilar
in their recursive calling patterns, and the hard work in the proof
involves demonstrating that both functions ultimately call a certain
(unknown) function on the same multiset of arguments (albeit in a
different order).  The clearest way to establish this fact is using
\emph{inductive} reasoning about computations on lists and trees,
which can be accomplished using standard proof techniques and is
orthogonal to the coinductive, relational style of reasoning that LSLR
(and in particular the $\later$ operator) provides.  While for this
example the inductive and coinductive bits of the proof can be
easily combined using a transitive combination of equivalences, it
would be interesting to explore in future work how to better integrate
inductive reasoning into our logic.

Bisimulations have also been developed for relational reasoning in
languages with general references and/or control
operators~\cite{Koutavas:Wand:06a,Stoevring-Lassen:POPL2007,sangiorgi-ea-2007,sumii:csl09}.
We hope that the present work will help to illuminate the relationship
between step-indexed logical relations and bisimulation techniques,
perhaps leading to a more unifying account.

Also related to our use of the \rul{l\"ob} rule is the work of Brandt and
Henglein~\cite{brandt-henglein}, who gave a coinductive axiomatization
of recursive type equality and subtyping via a coinduction-like
rule.  They also defined the semantic interpretation of their subtyping
judgment using a stratified, essentially step-indexed, interpretation.

Finally, besides step-indexed logical relations, a number of other
logical relations methods have been proposed for languages with
parametric polymorphism, recursion, and/or recursive types, \eg
\cite{pitts:parpoe,pitts:attapl,johann:voigtlaender,mellies-vouillon,birkedal,crary-harper-2007}.
One of the most important advances in this domain is the idea of
$\top\top$-closure (aka \emph{biorthogonality}).  In developing a
logical relation for a language with impredicative polymorphism,
existential types, and general recursion,
Pitts~\cite{pitts:parpoe,pitts:attapl} proposed $\top\top$-closure as
a useful operational technique for guaranteeing admissibility of
relations (in the denotational sense).  In the step-indexed model, the
whole issue of admissibility is sidestepped.  Intuitively, there is no
need to worry about a fixed-point behaving like the limit of its
finite approximations if we restrict attention to how programs behave
in a finite amount of time (as the step-indexed model does).

For non-step-indexed logical relations it is well-known that
$\top\top$-closure also has the pleasing side effect of rendering the
relations complete w.r.t. contextual equivalence.  This is also the
case for step-indexed logical relations, as shown in recent work of
Dreyer~\etal~\cite{dreyer+:icfp10}.  We have presented in this paper
an alternative technique for ensuring completeness, namely closure
w.r.t.\ $ciu$-approximation (in the definition of
$\Crel{\tau}{\rho}$).  We believe our approach is simpler and more
direct than $\top\top$-closure, but neither approach subsumes the
either.  On the one hand, $\top\top$-closure is applicable in more
general settings, such as lower-level
languages~\cite{benton-hur09,hur-dreyer11} or languages with control
operators~\cite{dreyer+:icfp10}, where the behavior of a term depends
on its evaluation context.  On the other hand, this added generality
means that a $\top\top$-closed relation is incapable of validating
some of the inference rules that hold in our more restricted setting.
For example, the \rul{sym-bind} rule (Figure~\ref{fig:symmetric})
would not hold in a $\top\top$-closed model unless we were to remove
the assumptions in the last premise connecting the $x_i$'s and the
$e_i$'s, thus weakening the rule somewhat.  We do believe, however,
that it should be possible to formalize a variant of our LSLR logical
relation that uses $\top\top$-closure instead of ciu-closure.
Understanding the tradeoffs between the two closure techniques remains
an interesting problem for future work.

Non-step-indexed logical relations for languages with \emph{recursive
  types} are notoriously tricky to construct; the construction of such
relations relies on the use of \emph{syntactic minimal invariance},
mimicking the construction used in domain
theory~\cite{pitts-reldom,birkedal,crary-harper-2007}.  An advantage
of this more elaborate construction over step-indexed logical
relations is that the resulting proof method is more abstract and does
not involve steps.  In this paper, we have shown how to devise a more
abstract proof method for step-indexed logical relations. Our
resulting proof method is at roughly the same level of abstraction as
that of non-step-indexed logical relations.  This point was
illustrated explicitly with the various examples in
Section~\ref{sec:examples}. For yet another example, just
involving recursive types, the reader might want to consider Birkedal
and Harper's example of stream operations~\cite{birkedal}.  Their
proof uses a coinduction proof principle that is derived as a
corollary of the elaborate construction of the logical relation.  This
example can also be proved in LSLR in a very similar manner, except
that we use a combination of the \rul{l\"ob} and
\rul{sym-exp-$\later$} rules instead of actual coinduction.

We do not claim that the method presented in this paper is \emph{per
  se} more powerful than prior approaches.  Rather, our goal is to
show how to reason about \emph{step-indexed} logical relations in a
more abstract way, because step-indexed relations have proven more
easily adaptable than other logical-relations methods to languages
with effects (particularly
state)~\cite{acar-ahmed-blume-2008,adr-popl09,neis+:icfp09}.  We
believe that the work presented here makes an important first step
toward \emph{logical} step-indexed logical relations for effectful
programs.  Indeed, since publication of our original LICS
paper~\cite{dreyer+:lics09}, a promising variant/extension of LSLR
(called LADR) has been developed~\cite{dreyer+:popl10}, which enables
abstract relational reasoning about a step-indexed model of $\fref$
(an extension of $\fmu$ with general references).

%%% Local Variables: 
%%% mode: latex
%%% TeX-master: "main"
%%% End: 

%% file: appendix.tex
\section{Additional Details of $\fmu$}
\label{sec:apdx:lang}

\newcommand{\AFigLangStaticSemII}[1][t!]{
\begin{figure}[#1]
\begin{myfig}
\vspace{1ex}
\begin{sdisplaymath}
  \begin{array}{l@{\quad}r@{\quad}c@{\quad}l}
\mbox{\textit{Typing Contexts}} &
\Gamma & \bnfdef & \empctx \bnfalt \Gamma, \alpha \bnfalt
\Gamma,x:\tau\\ 
\end{array}
\end{sdisplaymath}

\vspace{1ex}

\begin{flushleft}
\fbox{\small$\judg{\Gamma}{e}{\tau}$}
\end{flushleft}

\begin{smathpar}
\inferrule
{\Gamma(x) = \tau}
{\judg{\Gamma}{x}{\tau}}
\and
\inferrule
{ }
{\judg{\Gamma}{\eunit}{\tunit}}
\and
\inferrule
{ }
{\judg{\Gamma}{\pm n}{\tint}}
\\
\inferrule
{ }
{\judg{\Gamma}{\etrue}{\tbool}}
\and
\inferrule
{ }
{\judg{\Gamma}{\efalse}{\tbool}}
\and
\inferrule
{\judg{\Gamma}{e}{\tbool} 
 \\ \judg{\Gamma}{e_1}{\tau}
 \\ \judg{\Gamma}{e_2}{\tau}}
{\judg{\Gamma}{\eif{e}{e_1}{e_2}}{\tau}}
\\
\inferrule
{\judg{\Gamma}{e_1}{\tau_1}
 \\ \judg{\Gamma}{e_2}{\tau_2}}
{\judg{\Gamma}{\epair{e_1}{e_2}}{\tpair{\tau_1}{\tau_2}}}
\and
\inferrule
{\judg{\Gamma}{e}{\tpair{\tau_1}{\tau_2}}}
{\judg{\Gamma}{\efst{e}}{\tau_1}}
\and
\inferrule
{\judg{\Gamma}{e}{\tpair{\tau_1}{\tau_2}}}
{\judg{\Gamma}{\esnd{e}}{\tau_2}}
\\
\inferrule
{\judg{\Gamma}{e}{\tau_1}}
{\judg{\Gamma}{\einlty{\tsum{\tau_1}{\tau_2}}{e}}{\tsum{\tau_1}{\tau_2}}}
\and
\inferrule
{\judg{\Gamma}{e}{\tau_2}}
{\judg{\Gamma}{\einrty{\tsum{\tau_1}{\tau_2}}{e}}{\tsum{\tau_1}{\tau_2}}}
\and
\inferrule
{\judg{\Gamma}{e}{\tsum{\tau_1}{\tau_2}}
 \\ \judg{\Gamma,x_1:\tau_1}{e_1}{\tau} 
 \\ \judg{\Gamma,x_2:\tau_2}{e_2}{\tau}} 
{\judg{\Gamma}{\ecase{e}{x_1}{e_1}{x_2}{e_2}}{\tau}}
\\
\inferrule
{\judg{\Gamma,x:\tau_1}{e}{\tau_2}}
{\judg{\Gamma}{\efun{x}{\tau_1}{e}}{\tfun{\tau_1}{\tau_2}}}
\and
\inferrule
{\judg{\Gamma}{e_1}{\tfun{\tau_2}{\tau}} 
 \\ \judg{\Gamma}{e_2}{\tau_2}} 
{\judg{\Gamma}{\eapp{e_1}{e_2}}{\tau}}
\\
\inferrule
{\judg{\Gamma,\alpha}{e}{\tau}}
{\judg{\Gamma}{\etabs{\alpha}{e}}{\tall{\alpha}{\tau}}}
\and
\inferrule
{\judg{\Gamma}{e}{\tall{\alpha}{\tau}} 
 \\ \twfok{\Gamma}{\tau_1}} 
{\judg{\Gamma}{\etapp{e}{\tau_1}}{\subst{\tau_1}{\alpha}{\tau}}}
\\
\inferrule
{\twfok{\Gamma}{\tau_1} 
 \\ \judg{\Gamma}{e}{\subst{\tau_1}{\alpha}{\tau}}}
{\judg{\Gamma}{\epack{\tau_1}{e}{\texist{\alpha}{\tau}}}
       {\texist{\alpha}{\tau}}}
\and
\inferrule
{\judg{\Gamma}{e_1}{\texist{\alpha}{\tau_1}}
 \\ \twfok{\Gamma}{\tau} 
 \\ \judg{\Gamma,\alpha,x:\tau_1}{e_2}{\tau}}
{\judg{\Gamma}{\eunpack{e_1}{\alpha}{x}{e_2}}{\tau}}
\\
\inferrule
{\judg{\Gamma}{e}
       {\subst{\trec{\alpha}{\tau}}{\alpha}{\tau}} 
 \\ \twfok{\Gamma}{\trec{\alpha}{\tau}}}
{\judg{\Gamma}{\efoldty{\trec{\alpha}{\tau}}{e}}{\trec{\alpha}{\tau}}}
\and
\inferrule
{\judg{\Gamma}{e}{\trec{\alpha}{\tau}}}
{\judg{\Gamma}{\eunfold{e}}
       {\subst{\trec{\alpha}{\tau}}{\alpha}{\tau}}}
\end{smathpar}
%\vspace{-3ex}
\caption{$\fmu$ Static Semantics}
\label{fig:apdx:staticsem}
\end{myfig}
\end{figure}
}

\AFigLangStaticSemII[h]

\clearpage
\newcommand{\AFigLangContextsI}[1][th]{
\begin{figure*}[#1]
\begin{myfig}
\begin{sdisplaymath}
\begin{array}{l@{\quad}r@{\quad}c@{\quad}l}
\mbox{\textit{Contexts}} &
\ctxt & \bnfdef & \hole \bnfalt
\op(e_1,\ldots,e_{i-1},\ctxt,e_{i+1},\ldots,e_n)
\bnfalt 
\\ & & & 
\eif{\ctxt}{e_1}{e_2} \bnfalt 
\eif{e}{C}{e_2} \bnfalt 
\eif{e}{e_1}{C} \bnfalt 
\\ & & & 
\epair{\ctxt}{e_2} \bnfalt \epair{e_1}{\ctxt} \bnfalt 
\efst{\ctxt} \bnfalt \esnd{\ctxt} \bnfalt 
\\ & & & 
\einlty{\tau}{\ctxt} \bnfalt \einrty{\tau}{\ctxt} \bnfalt
\ecase{\ctxt}{x_1}{e_1}{x_2}{e_2} \bnfalt  
\\ & & & 
\ecase{e}{x_1}{\ctxt}{x_2}{e_2} \bnfalt  
\ecase{e}{x_1}{e_1}{x_2}{\ctxt} \bnfalt  
\\ & & & 
\efun{x}{\tau}{\ctxt} \bnfalt
\eapp{\ctxt}{e} \bnfalt \eapp{e}{\ctxt} \bnfalt 
\etabs{\alpha}{\ctxt} \bnfalt \etapp{\ctxt}{\tau} \bnfalt 
\\ & & & 
\epack{\tau_1}{\ctxt}{\texist {\alpha}{\tau}} \bnfalt
\eunpack{\ctxt}{\alpha}{x}{e_2} \bnfalt 
\eunpack{e_1}{\alpha}{x}{\ctxt} \bnfalt 
\\ & & & 
\efoldty{\tau}{\ctxt} \bnfalt
\eunfold{\ctxt} \bnfalt
\end{array}
\end{sdisplaymath}

\vspace{1ex}

\begin{flushleft}
\fbox{\small$\vdash \ctxt : (\twf{\Gamma}{\tau}) 
\ctxarrow (\twf{\Gamma'}{\tau'}) 
$}
\end{flushleft}

\begin{smathpar}
\inferrule
{\Gamma \subseteq \Gamma'}
{\vdash \hole : (\twf{\Gamma}{\tau}) \ctxarrow 
                (\twf{\Gamma'}{\tau})}
\and
\inferrule
{\vdash C : (\twf{\Gamma}{\tau}) \ctxarrow 
            (\twf{\Gamma'}{\tbool}) 
 \\ \judg{\Gamma'}{e_1}{\tau'}
 \\ \judg{\Gamma'}{e_2}{\tau'}}
{\vdash \eif{C}{e_1}{e_2} : (\twf{\Gamma}{\tau}) \ctxarrow 
                (\twf{\Gamma'}{\tau'})}
\and
\inferrule
{\judg{\Gamma'}{e}{\tbool} 
 \\ \vdash C : (\twf{\Gamma}{\tau}) \ctxarrow 
            (\twf{\Gamma'}{\tau'}) 
 \\ \judg{\Gamma'}{e_2}{\tau'}}
{\vdash \eif{e}{C}{e_2} : (\twf{\Gamma}{\tau}) \ctxarrow 
                (\twf{\Gamma'}{\tau'})}
\and
\inferrule
{\judg{\Gamma'}{e}{\tbool} 
 \\ \judg{\Gamma'}{e_1}{\tau'}
 \\ \vdash C : (\twf{\Gamma}{\tau}) \ctxarrow 
            (\twf{\Gamma'}{\tau'})}
{\vdash \eif{e}{e_1}{C} : (\twf{\Gamma}{\tau}) \ctxarrow 
                (\twf{\Gamma'}{\tau'})}
\and
\inferrule
{\vdash C : (\twf{\Gamma}{\tau}) \ctxarrow 
            (\twf{\Gamma'}{\tau_1})
 \\ \judg{\Gamma'}{e_2}{\tau_2}}
{\vdash \epair{C}{e_2} : (\twf{\Gamma}{\tau}) \ctxarrow 
                (\twf{\Gamma'}{\tpair{\tau_1}{\tau_2}})}
\and
\inferrule
{\judg{\Gamma'}{e_1}{\tau_1}
 \\ \vdash C : (\twf{\Gamma}{\tau}) \ctxarrow 
            (\twf{\Gamma'}{\tau_2})}
{\vdash \epair{e_1}{C} : (\twf{\Gamma}{\tau}) \ctxarrow 
                (\twf{\Gamma'}{\tpair{\tau_1}{\tau_2}})}
\\
\inferrule
{\vdash C : (\twf{\Gamma}{\tau}) \ctxarrow 
            (\twf{\Gamma'}{\tpair{\tau_1}{\tau_2}})}
{\vdash \efst{C} : (\twf{\Gamma}{\tau}) \ctxarrow 
            (\twf{\Gamma'}{\tau_1})}
\and
\inferrule
{\vdash C : (\twf{\Gamma}{\tau}) \ctxarrow 
            (\twf{\Gamma'}{\tpair{\tau_1}{\tau_2}})}
{\vdash \esnd{C} : (\twf{\Gamma}{\tau}) \ctxarrow 
            (\twf{\Gamma'}{\tau_2})}
\\
\inferrule
{\vdash C : (\twf{\Gamma}{\tau}) \ctxarrow 
            (\twf{\Gamma'}{\tau_1})}
{\vdash \einlty{\tsum{\tau_1}{\tau_2}}{C} : (\twf{\Gamma}{\tau}) \ctxarrow 
            (\twf{\Gamma'}{\tsum{\tau_1}{\tau_2}})}
\and
\inferrule
{\vdash C : (\twf{\Gamma}{\tau}) \ctxarrow 
            (\twf{\Gamma'}{\tau_2})}
{\vdash \einrty{\tsum{\tau_1}{\tau_2}}{C} : (\twf{\Gamma}{\tau}) \ctxarrow 
            (\twf{\Gamma'}{\tsum{\tau_1}{\tau_2}})}
\and
\inferrule
{\vdash C : (\twf{\Gamma}{\tau}) \ctxarrow 
            (\twf{\Gamma'}{\tsum{\tau_1}{\tau_2}})
 \\ \judg{\Gamma',x_1:\tau_1}{e_1}{\tau'} 
 \\ \judg{\Gamma',x_2:\tau_2}{e_2}{\tau'}} 
{\vdash \ecase{C}{x_1}{e_1}{x_2}{e_2} : 
        (\twf{\Gamma}{\tau}) \ctxarrow 
        (\twf{\Gamma'}{\tau'})}
\\
\inferrule
{\judg{\Gamma'}{e}{\tsum{\tau_1}{\tau_2}} 
 \\ \vdash C : (\twf{\Gamma}{\tau}) \ctxarrow 
            (\twf{\Gamma',x_1:\tau_1}{\tau'})
 \\ \judg{\Gamma',x_2:\tau_2}{e_2}{\tau'}} 
{\vdash \ecase{e}{x_1}{C}{x_2}{e_2} : 
        (\twf{\Gamma}{\tau}) \ctxarrow 
        (\twf{\Gamma'}{\tau'})}
\and
\inferrule
{\judg{\Gamma'}{e}{\tsum{\tau_1}{\tau_2}} 
 \\ \judg{\Gamma',x_1:\tau_1}{e_1}{\tau'}
 \\ \vdash C : (\twf{\Gamma}{\tau}) \ctxarrow 
            (\twf{\Gamma',x_2:\tau_2}{\tau'})}
{\vdash \ecase{e}{x_1}{e_1}{x_2}{C} : 
        (\twf{\Gamma}{\tau}) \ctxarrow 
        (\twf{\Gamma'}{\tau'})}
\\
\inferrule
{\vdash C : (\twf{\Gamma}{\tau}) \ctxarrow 
        (\twf{\Gamma',x:\tau_1}{\tau_2})}
{\vdash \efun{x}{\tau_1}{C} : 
        (\twf{\Gamma}{\tau}) \ctxarrow 
        (\twf{\Gamma'}{\tfun{\tau_1}{\tau_2}})}
\and
\inferrule
{\vdash C : (\twf{\Gamma}{\tau}) \ctxarrow 
            (\twf{\Gamma'}{\tfun{\tau_2}{\tau'}}) 
 \\ \judg{\Gamma'}{e_2}{\tau_2}} 
{\vdash \eapp{C}{e_2} : (\twf{\Gamma}{\tau}) \ctxarrow 
        (\twf{\Gamma'}{\tau'})}
\and
\inferrule
{\judg{\Gamma'}{e_1}{\tfun{\tau_2}{\tau'}}
 \\ \vdash C : (\twf{\Gamma}{\tau}) \ctxarrow 
            (\twf{\Gamma'}{\tau_2})}
{\vdash \eapp{e_1}{C} : (\twf{\Gamma}{\tau}) \ctxarrow 
        (\twf{\Gamma'}{\tau'})}
\end{smathpar}
\caption{$\fmu$ Program Contexts: Syntax and Static Semantics I}
\label{fig:apdx:contextsI}
\end{myfig}
\end{figure*}
}

\AFigLangContextsI[h!]

\newcommand{\AFigLangContextsII}[1][th]{
\begin{figure*}[#1]
\begin{myfig}

\begin{flushleft}
\fbox{\small$\vdash \ctxt : (\twf{\Gamma}{\tau}) 
\ctxarrow (\twf{\Gamma'}{\tau'}) 
$} \hspace{0.1in}\small{(contd. from Figure~\ref{fig:apdx:contextsI})}
\end{flushleft}
\begin{smathpar}
\inferrule
{\vdash C : (\twf{\Gamma}{\tau}) \ctxarrow 
        (\twf{\Gamma',\alpha}{\tau'})}
{\vdash \etabs{\alpha}{C} : (\twf{\Gamma}{\tau}) \ctxarrow 
        (\twf{\Gamma'}{\tall{\alpha}{\tau'}})}
\and
\inferrule
{\vdash C : (\twf{\Gamma}{\tau}) \ctxarrow 
        (\twf{\Gamma'}{\tall{\alpha}{\tau'}}) 
 \\ \twfok{\Gamma'}{\tau_1}} 
{\vdash \etapp{C}{\tau_1} : (\twf{\Gamma}{\tau}) \ctxarrow 
        (\twf{\Gamma'}{\subst{\tau_1}{\alpha}{\tau'}})}
\\
\inferrule
{\twfok{\Gamma'}{\tau_1} 
 \\ \vdash C : (\twf{\Gamma}{\tau}) \ctxarrow 
        (\twf{\Gamma'}{\subst{\tau_1}{\alpha}{\tau'}})}
{\vdash \epack{\tau_1}{C}{\texist{\alpha}{\tau'}} : 
           (\twf{\Gamma}{\tau}) \ctxarrow 
           (\twf{\Gamma'}{\texist{\alpha}{\tau'}})}
\and
\inferrule
{\vdash C : (\twf{\Gamma}{\tau}) \ctxarrow 
        (\twf{\Gamma'}{\texist{\alpha}{\tau_1}})
 \\ \twfok{\Gamma'}{\tau'} 
 \\ \judg{\Gamma',\alpha,x:\tau_1}{e_2}{\tau'}}
{\vdash \eunpack{C}{\alpha}{x}{e_2} : 
        (\twf{\Gamma}{\tau}) \ctxarrow 
        (\twf{\Gamma'}{\tau'})}
\and
\inferrule
{\judg{\Gamma'}{e_1}{\texist{\alpha}{\tau_1}}
 \\ \twfok{\Gamma'}{\tau'} 
 \\ \vdash C : (\twf{\Gamma}{\tau}) \ctxarrow 
        (\twf{\Gamma',\alpha,x:\tau_1}{\tau'})}
{\vdash \eunpack{e_1}{\alpha}{x}{C} : 
        (\twf{\Gamma}{\tau}) \ctxarrow 
        (\twf{\Gamma'}{\tau'})}
\\
\inferrule
{\vdash C : (\twf{\Gamma}{\tau}) \ctxarrow 
        (\twf{\Gamma'}
             {\subst{\trec{\alpha}{\tau'}}{\alpha}{\tau'}}) 
 \quad \twfok{\Gamma'}{\trec{\alpha}{\tau'}}}
{\vdash \efoldty{\trec{\alpha}{\tau'}}{C} : (\twf{\Gamma}{\tau}) \ctxarrow 
        (\twf{\Gamma'}{\trec{\alpha}{\tau'}})}
\hfill
\inferrule
{\vdash C : (\twf{\Gamma}{\tau}) \ctxarrow 
        (\twf{\Gamma'}
             {\trec{\alpha}{\tau'}})}
{\vdash \eunfold{C} : (\twf{\Gamma}{\tau}) \ctxarrow 
        (\twf{\Gamma'}
             {\subst{\trec{\alpha}{\tau'}}{\alpha}{\tau'}})}
\end{smathpar}
\caption{$\fmu$ Program Contexts: Static Semantics II}
\vspace{0.3in}
\label{fig:apdx:contextsII}
\end{myfig}
\end{figure*}
}

\AFigLangContextsII[t]

%%% Local Variables: 
%%% mode: latex
%%% TeX-master: "main"
%%% End: 

%% file: lang-stepless.tex
\mbox{}
\vspace{-5ex}
\section{Remaining Inference Rules for \thelogic{}}
\label{sec:apdx:logic}

\noindent
Here, we present the LSLR judgments of relation and substitution
well-formedness, as well as additional inference rules that are entirely standard.
$\Prop$ is
synonymous with $\Rel(0)$.

\newcommand{\AFigRulesLSLRI}[1][th]{
\begin{flushleft}
\fbox{\small$\gr \ts R :: \Rel(n)$} 
\end{flushleft}
\begin{smathpar}
\inferrule{
r \in \rcon \+
\arity(r) = n
}{
\gr \ts r :: \Rel(n)
}
\and
\inferrule{
\fv(e_1,e_2) \subseteq \vcon
}{
\gr \ts e_1 = e_2 :: \Prop
}
\and
\inferrule{
}{
\gr\ts \OVal :: \Rel(1)
}
\and
\inferrule{
\fv(e,\tau) \subseteq \vcon
}{
\gr\ts e : \tau :: \Prop
}
\and
\inferrule{
\fv(C,\tau,\tau') \subseteq \vcon
% \quad \bv(C) = \emptyset
}{
\gr\ts C : \tau \eto \tau' :: \Prop
}
\and
\inferrule{
\fv(e_1,e_2) \subseteq \vcon
}{
\gr\ts e_1 \step^{\{*,0,1\}} e_2 :: \Prop
}
\and
\inferrule{
\fv(e_1,e_2) \subseteq \vcon
}{
\gr\ts e_1 \ciuleq e_2 :: \Prop
}
\\
\inferrule{
}{
\gr \ts \top :: \Prop
}
\and
\inferrule{
}{
\gr \ts \bot :: \Prop
}
\and
\inferrule{
\gr \ts P :: \Prop \+
\gr \ts Q :: \Prop
}{
\gr \ts P \land Q :: \Prop
}
\and
\inferrule{
\gr \ts P :: \Prop \+
\gr \ts Q :: \Prop
}{
\gr \ts P \lor Q :: \Prop
}
\and
\inferrule{
\gr \ts P :: \Prop \+
\gr \ts Q :: \Prop
}{
\gr \ts P \limp Q :: \Prop
}
\\
\inferrule{
\vcon,\vcon';\rcon \ts P :: \Prop
}{
\gr \ts \forall\vcon'.P :: \Prop
}
\hfill%qquad
\inferrule{
\vcon;\rcon,\rcon' \ts P :: \Prop
}{
\gr \ts \forall\rcon'.P :: \Prop
}
\hfill%qquad
\inferrule{
\vcon,\vcon';\rcon \ts P :: \Prop
}{
\gr \ts \exists\vcon'.P :: \Prop
}
\hfill%qquad
\inferrule{
\vcon;\rcon,\rcon' \ts P :: \Prop
}{
\gr \ts \exists\rcon'.P :: \Prop
}
\\
\inferrule{
\vcon,\seq{x};\rcon \ts P :: \Prop \+
\seq{x} = x_1,\ldots,x_n
}{
\gr \ts \seq{x}.P :: \Rel(n)
}
\and
\inferrule{
\fv(\seq{e}) \subseteq \vcon \+
\seq{e} = e_1,\ldots,e_n \+
\gr \ts R :: \Rel(n)
}{
\gr \ts \seq{e} \in R :: \Prop
}
\and
\and
\inferrule{
\gr, r \ts R :: \Rel(n) \+
\arity(r) = n \+
\mbox{$R$ contractive in $r$}
}{
\gr \ts \mu r. R :: \Rel(n)
}
\and
\inferrule{
\gr \ts P :: \Prop
}{
\gr \ts \lift P :: \Prop
}
\end{smathpar}
}

\AFigRulesLSLRI[h!]

\newcommand{\AFigRulesLSLRII}[1][th]{
\begin{flushleft}
\fbox{\small$\vcon \ts \gamma :: \vcon' 
$} 
\end{flushleft}
\begin{smathpar}
\inferrule{
\dom(\gamma) = \vcon' \+
\forall \alpha \in \vcon'.\ \fv(\gamma\alpha) \subseteq \vcon \+
\forall x \in \vcon'.\ \fv(\gamma x) \subseteq \vcon
}{
\vcon \ts \gamma :: \vcon'
}
\end{smathpar}

%%%%
\begin{flushleft}
\fbox{\small$\gr \ts \varphi :: \rcon' 
$} 
\end{flushleft}
\begin{smathpar}
\inferrule{
\dom(\varphi) = \rcon' \+
\forall r \in \rcon'.\ \arity(r) = n \limp \gr \ts \varphi r :: \Rel(n)
}{
\gr \ts \varphi :: \rcon'
}
\end{smathpar}

%%%%%%%%%%%%%%%%%%%
\begin{flushleft}
\fbox{\small$\gra \ts P
$} 
\end{flushleft}
\begin{smathpar}
\inferrule{
\gra \ts P
}{
\gra,\gra' \ts P
}
\and
\inferrule{
}{
\gra,P \ts P
}
\and
\inferrule{
}{
\gra \ts \top
}
\and
\inferrule{
\gra \ts \bot
}{
\gra \ts P
}
\and
\inferrule{
\gra \ts P \quad
\gra \ts Q
}{
\gra \ts P \land Q
}
\and
\inferrule{
\gra \ts P \land Q
}{
\gra \ts P
}
\and
\inferrule{
\gra \ts P \land Q
}{
\gra \ts Q
}
\\
\inferrule{
\gra \ts P
}{
\gra \ts P \lor Q
}
\and
\inferrule{
\gra \ts Q
}{
\gra \ts P \lor Q
}
\and
\inferrule{
\gra \ts P \lor Q \+
\gra,P \ts C \+
\gra,Q \ts C
}{
\gra \ts C
}
\\
\inferrule{
\gra,P \ts Q
}{
\gra \ts P \limp Q
}
\and
\inferrule{
\gra \ts P \limp Q \+
\gra \ts P
}{
\gra \ts Q
}
\\
\inferrule{
\gra,\vcon' \ts P
}{
\gra \ts \forall\vcon'.P
}
\and
\inferrule{
\gra \ts \forall\vcon'.P \+
\gra \ts \gamma :: \vcon'
}{
\gra \ts \gamma P
}
\and
\inferrule{
\gra,\rcon' \ts P
}{
\gra \ts \forall\rcon'.P
}
\and
\inferrule{
\gra \ts \forall\rcon'.P \+
\gra \ts \varphi :: \rcon'
}{
\gra \ts \varphi P
}
\and
\inferrule{
\gra \ts \gamma :: \vcon' \+
\gra \ts \gamma P
}{
\gra \ts \exists\vcon'.P
}
\and
\inferrule{
\gra \ts \exists\vcon'.P \+
\gra,\vcon',P \ts Q
}{
\gra \ts Q
}
\\
\inferrule{
\gra \ts \varphi :: \rcon' \+
\gra \ts \varphi P
}{
\gra \ts \exists\rcon'.P
}
\and
\inferrule{
\gra \ts \exists\rcon'.P \+
\gra,\rcon',P \ts Q
}{
\gra \ts Q
}

\end{smathpar}
}

\AFigRulesLSLRII[h!]

%%% Local Variables: 
%%% mode: latex
%%% TeX-master: "main"
%%% End: 